\documentclass[fleqn,usenatbib]{mnras}

\usepackage{newtxtext,newtxmath}

\usepackage[T1]{fontenc}

\DeclareRobustCommand{\VAN}[3]{#2}
\let\VANthebibliography\thebibliography
\def\thebibliography{\DeclareRobustCommand{\VAN}[3]{##3}\VANthebibliography}

\usepackage{graphicx}	
\usepackage{amsmath}	
\usepackage[dvipsnames]{xcolor}

\newcommand{\simname}[1]{\textit{#1}}

\usepackage{soul}

\title[Jet feedback in halos]{Active galactic nucleus jet feedback in hydrostatic halos}

\author[R. Weinberger et al.]{%
Rainer Weinberger,$^{1}$\thanks{E-mail: rainer@cita.utoronto.ca}
Kung-Yi Su,$^{2}$
Kristian Ehlert,$^{3,4}$
Christoph Pfrommer,$^{3}$
Lars Hernquist,$^{5}$
\newauthor
Greg L. Bryan,$^{6,7}$
Volker Springel,$^{8}$
Yuan Li,$^{9}$
Blakesley Burkhart,$^{7,10}$
Ena Choi,$^{11,12}$
\newauthor
Claude-Andr\'e Faucher-Gigu\`ere$^{13}$
\\
$^{1}$Canadian Institute for Theoretical Astrophysics, 60 St. George Street, Toronto, ON M5S 3H8, Canada\\
$^{2}$Black Hole Initiative, Harvard University, 20 Garden St., Cambridge, MA 02138, USA\\
$^{3}$Leibniz Institute for Astrophysics, An der Sternwarte 16, 14482 Potsdam, Germany\\
$^{4}$Institut f\"ur Physik und Astronomie, Universit\"at Potsdam, Karl-Liebknecht-Str. 24/25, 14476 Golm, Germany\\
$^{5}$Center for Astrophysics | Harvard \& Smithsonian, 60 Garden Street, Cambridge, MA 02138, USA \\
$^{6}$Department of Astronomy, Columbia University, 550 West 120th Street, New York, NY 10027, USA\\
$^{7}$Center for Computational Astrophysics, Flatiron Institute, 162 Fifth Avenue, New York, NY 10010, USA\\
$^{8}$Max-Planck-Institute for Astrophysics, Karl-Schwarzschild-Str. 1, D-85741 Garching, Germany \\
$^{9}$University of North Texas, 1155 Union Circle 311277 Denton, TX 76203-5017, USA  \\
$^{10}$Department of Physics and Astronomy, Rutgers University,
136 Frelinghuysen Rd, Piscataway, NJ 08854, USA \\
$^{11}$Department of Physics, University of Seoul, 163 Seoulsiripdaero, Dongdaemun-gu, Seoul 02504, Republic of Korea \\
$^{12}$Korea Institute for Advanced Study (KIAS), 85 Hoegiro, Dongdaemun-gu, Seoul 02455, Republic of Korea \\
$^{13}$Department of Physics and Astronomy and CIERA,
Northwestern University, 2145 Sheridan Road, Evanston, IL 60208, USA
}

\date{Accepted XXX. Received YYY; in original from ZZZ}

\pubyear{2023}

\begin{document}
\label{firstpage}
\pagerange{\pageref{firstpage}--\pageref{lastpage}}
\maketitle

\begin{abstract}
Feedback driven by jets from active galactic nuclei is believed to be responsible for reducing cooling flows in cool-core galaxy clusters.
We use simulations to model feedback from hydrodynamic jets in isolated halos.
While the jet propagation converges only after the diameter of the jet is well resolved, reliable predictions about the effects these jets have on the cooling time distribution function only require resolutions sufficient to keep the jet-inflated cavities stable.
Comparing different model variations, as well as an independent jet model using a different hydrodynamics code, we show that the dominant uncertainties are the choices of jet properties within a given model.
Independent of implementation, we find that light, thermal jets with low momentum flux tend to delay the onset of a cooling flow more efficiently on a $50$~Myr timescale than heavy, kinetic jets.
The delay of the cooling flow originates from a displacement and boost in entropy of the central gas.
If the jet kinetic luminosity depends on accretion rate, collimated, light, hydrodynamic jets are able to reduce cooling flows in halos, without a need for jet precession or wide opening angles.
Comparing the jet feedback with a `kinetic wind' implementation shows that equal amounts of star formation rate reduction can be achieved by different interactions with the halo gas: the jet has a larger effect on the hot halo gas while leaving the denser, star forming phase in place, while the wind acts more locally on the star forming phase, which manifests itself in different time-variability properties.
\end{abstract}




\section{Introduction}

Feedback from active galactic nuclei (AGNs) has proven to be essential in theoretical studies of galaxy formation.
It can be used as a mechanism to break the self-similarity of the stellar mass function \citep{bower06} and bring galaxy colors and the galaxy luminosity function at the massive end in agreement with observations \citep{croton06}.
State of the art cosmological simulations of galaxy formation rely on these feedback effects to reproduce the observed low rates of star formation in massive central galaxies \citep{vogelsberger14a,vogelsberger14b,genel14,somerville15, naab17} as well as the neutral gas content in the intergalactic medium \citep{tillman22}.
The implementations of these feedback effects operate at the resolution limit of these simulations, on $100$~pc to kpc scales.
AGN driven winds and jets, however, originate from spatial scales orders of magnitude smaller.
Even the most recent simulations designed for high dynamic range are not yet able to fully cover all scales from the central engine to energy deposition \citep{chatterjee19, lalakos22}.
This unclear outflow behaviour at kpc scales, combined with numerical uncertainties that increase close to the resolution scale, is a key factor currently limiting the predictive power of cosmological galaxy formation simulations.

Observationally, the most direct window into the process of AGN feedback can be found in massive cool-core galaxy clusters \citep{fabian12, mcnamara12}.
These clusters show remarkably low levels of star formation despite high observed X-ray luminosities, indicative of radiative cooling, which has to be balanced by a heating source for the hot gas in these systems to avoid overcooling and substantial star formation.
The jet-inflated X-ray dark cavities, ubiquitous in these systems, are the primary candidate for providing this heating \citep{churazov02, reynolds02}.
Using the approximate hydrostatic nature of the intra-cluster medium (ICM), it is possible to infer a heating rate from the sizes and locations of these cavities.
The inferred energy injection rate correlates remarkably well with the cooling luminosity of clusters and groups \citep{birzan04}, indicating that, energetically, jets are able to mediate cooling flows in galaxy groups and clusters and explain their low star formation rates.
Yet, the precise mechanism determining how this energy is spatially distributed and how gas is heated is still under debate.

Proposed mechanisms include mixing of the jet lobe material \citep{hillel17}, convective energy transport \citep{yang16, chen19}, turbulent dissipation \citep{fujita20}, shock heating \citep{li17}, sound and gravity waves \citep{reynolds15, bambic19}, cosmic ray protons \citep{guo08, jacob17, jacob17b, ruszkowski17}, thermal conduction \citep{yang16b, kannan17, barnes19}, as well as a combination of a number of the above mentioned effects \citep[e.g.][]{soker19, su20}.

One of the underlying problems at the heart of studying how AGN jet energy is distributed is the enormous dynamic range of scales involved.
This implies that either simulations cover the long-term evolution, modeling the system for several cooling times to ensure star formation is affected and a steady-state in the cooling flow is reached \citep[e.g.][]{dubois10, gaspari11, gaspari11b, gaspari12, prasad15, meece17, martizzi19, beckmann19, husko22c}, or simulations resolve jets and their propagation in more detail, but are limited to 1 to 2 orders of magnitude shorter simulation times. The latter simulations cover only a buoyancy timescale, or even only the jet propagation phase \citep[e.g.][]{bourne19, bourne20, duan20, komissarov21, talbot21, talbot22, massaglia21, husko22, perucho22}.
Our work is at the intersection of these two approaches, trying to link high fidelity jet propagation studies with studies of self-regulated cooling flows in hydrostatic halos in an overall attempt to create a predictive AGN jet feedback model. In particular, we investigate whether resolved hydrodynamic jets are able to moderate cooling flows in galaxy clusters over Gyr timescales and how this ability depends on numerical resolution, model choices and implementation/code details.

The paper is structured as follows:
we present the simulation methods and setup in Section~\ref{sec:model}, study the jet propagation of a single outburst in Section~\ref{sec:jet_propagation} and its effect on hydrostatic halos in Section~\ref{sec:effect_on_halo}. In Section~\ref{sec:self_regulated}, we present self-regulated simulations and discuss the impact of AGN jets on them. Finally, we present our conclusions from this study in Section~\ref{sec:conclusion}.

\section{Model and Simulation Setup}
\label{sec:model}

To get a sense of how AGN jets act in an ICM environment, we model an isolated hydrostatic halo of mass $10^{14}$~M$_\odot$ at $z=1$, using an analytic gravitational potential and perform simulations including radiative cooling, star formation, and AGN feedback.
In a first set of simulations, we inject AGN feedback at fixed luminosity $\dot{E} = 10^{45}$~erg~s$^{-1}$ using two independent jet model implementations and different simulation codes, Arepo \citep{springel10} and Gizmo \citep{hopkins15}, as well as an AGN wind model.
These simulations only cover the onset of a cooling flow, and do not end up forming any stars in their $250$~Myr simulation time.
A second set of simulations uses the same initial conditions, but now the black hole accretion rate is estimated using the Bondi rate and $10$ per cent of the accreted rest mass energy is used as feedback energy.
These runs are evolved for several central cooling times (in total $2$~Gyr), to ensure that self-regulation actually sets in, and to be able to study the effects on the star formation rate.

\subsection{Initial Conditions}

\begin{table}
    \centering
    \begin{tabular}{lc}
    \hline
    Parameter & Value \\
    \hline
         M$_\text{NFW}$ & $10^{14} \text{ M}_\odot$ \\
         R$_\text{NFW}$ & $1.0$ \text{ Mpc} \\
         $c_\text{NFW}$ & $2.84$ \\
         M$_\text{ISO}$ & $3.7 \times 10^9 \text{ M}_\odot$ \\
         R$_\text{ISO}$ & $0.1 \text{ Mpc}$ \\
         \hline
         $\rho_1$ & $1.13 \times 10^{-25} \text{g cm}^{-3}$ \\
         $r_1$ & $26.3 \text{ kpc}$ \\
         $\beta_1$ & $1.89$ \\
         $\rho_2$ & $7.16 \times 10^{-27} \text{g cm}^{-3}$ \\
         $r_2$ & $303 \text{ kpc}$ \\
         $\beta_2$ & $1.42$ \\
         \hline
         $\sigma_\text{gas}$ & $75 \text{ km s}^{-1}$ \\
         $k_\text{min}$ & $8.4 \text{ Mpc}^{-1}$\\
         $k_\text{max}$ & $16.8 \text{ Mpc}^{-1}$ \\
         \hline
    \end{tabular}
    \caption{Parameters for the initial conditions. The last three rows specify the turbulent power spectrum of the initial gas velocities. The resulting cooling luminosity of the halo gas is approximately $10^{44}$~erg~s$^{-1}$.}
    \label{tab:ics}
\end{table}

\begin{figure}
    \centering
    \includegraphics{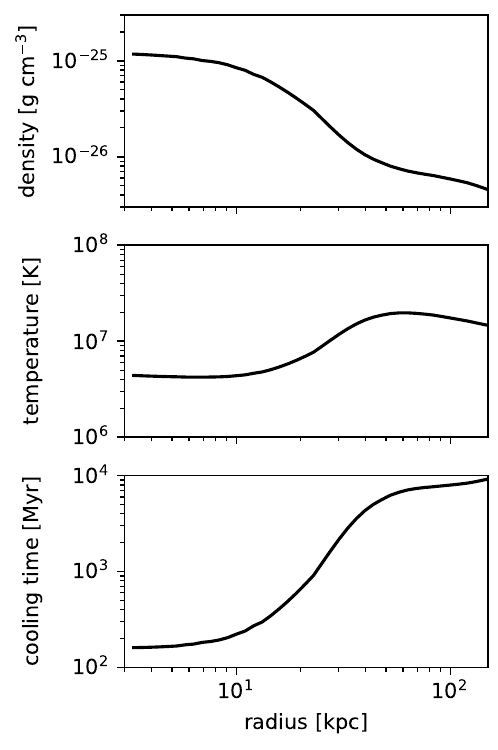}
    \caption{Density, temperature and cooling time profiles of initial conditions.}
    \label{fig:ic_profiles}
\end{figure}

Unlike a number of similar studies which focus on matching their model to local galaxy clusters, we choose to base our initial conditions on cosmological simulations of galaxy clusters at redshift $z=1$. This is done for a number of reasons: first, starting with a local cluster implies simulating into the future and second, the ICM conditions can change with redshift. Concretely, high redshift galaxy clusters are more frequently found to be in a cool-core state \citep{mcdonald13, mcdonald17,barnes19} with higher cooling luminosities and AGN powers \citep{weinberger18}, making the cooling-flows more prominent. Considering that our assumption of an isolated hydrostatic halo degrades with increasing redshift, we choose redshift $z=1$ as a compromise. Note that the simulation itself neglects cosmological expansion and all length scales are stated in proper coordinates.

The hydrostatic halo in this study is a fit to $z=1$ halos in the IllustrisTNG cosmological simulations \citep{marinacci18, naiman18, nelson18, pillepich18b, springel18}.
We use a truncated NFW profile \citep{navarro96} and singular isothermal sphere profile to fit the enclosed gravitating mass:

\begin{align}
 M_\text{enc}(r) &= \min\left\{4 \pi \rho_\text{NFW} R_\mathrm{s}^3 \left[\ln\left(\frac{r + R_\mathrm{s}}{R_\mathrm{s}}\right) - \frac{r}{r+R_\mathrm{s}}\right], M_\text{NFW}\right\} \nonumber\\
  &+ \min\left\{\frac{r \,M_\text{ISO}}{R_\text{ISO}}, M_\text{ISO}\right\},
\end{align}
where
\begin{align}
\rho_\text{NFW} &= M_\text{NFW} \left[ 4\pi\, R_\mathrm{s}^3\,  \left(\ln(1+c_\text{NFW})-\frac{c_\text{NFW}}{1+c_\text{NFW}}\right)\right]^{-1},\label{eq:rhoNFW}\\
R_\mathrm{s} &= R_\text{NFW}\, c_\text{NFW}^{-1}.
\end{align}
Note that we fixed $R_\text{ISO} = 0.1$~\text{ Mpc} to avoid degeneracies in the fit.

The gas density profile is fitted with a double beta profile:
\begin{align}
    \rho(r) = \left(\frac{\rho_1}{1 + (r/r_1)^2}\right)^{3/2 \,\beta_1} + \left(\frac{\rho_2}{1 + (r/r_2)^2}\right)^{3/2 \,\beta_2}
\end{align}

The thermal pressure is adjusted such that the gas halo is initially in hydrostatic equilibrium.
To break the exact spherical symmetry, the gas velocities in the central region ($r<600$~kpc) follow a Kolmogorov power spectrum, normalized to a root mean square velocity of $75$~km~s$^{-1}$.\footnote{This is lower than the measured velocity dispersion in local massive galaxy clusters \citep{zhuravleva2014, hitomi2016, li2020}.}
Gas further out is initially at rest. We place a black hole at rest in the center of the potential, and keep it fixed at this position during the simulation.
All the parameters related to the initial conditions are shown in Table~\ref{tab:ics} and the profiles of density, temperature and cooling time ($t_\text{c} = -\varepsilon_\text{th}\, \dot{\varepsilon}^{-1}_\text{c}$ where $\varepsilon_\text{th}$ is the thermal energy density and $\dot{\varepsilon}^{-1}_\text{c}$ its rate of change due to radiative cooling) are indicated in Fig.~\ref{fig:ic_profiles}.

The discretization of the hydrodynamic quantities follows the approach of \citet{pakmor11, ohlmann16} and ensures that both volume-discretized schemes such as Arepo and mass discretized schemes such as employed in the meshless finite mass scheme used in Gizmo can handle the same input.
We employ a radius dependent mass per resolution element $i$ following

\begin{align}
    m_i = m_0 \exp\left(\frac{r}{r_0}\right),
\end{align}
with $m_0=9.4\times 10^4\,\text{M}_\odot$ and $r_0=3\times 10^2$~kpc, but limiting the volume of a resolution element (defined as its mass divided by its density) to a maximum of $5\times 10^7$~kpc$^3$. The latter is only there to avoid numerical problems in the outskirts and has no impact on the regions studied in this work. The overall simulation domain has the side length $6\times 10^3$~kpc which ensures that the region of interest is free of boundary effects.
During the simulation, cells are refined and derefined to stay within a factor of 2 of the target mass $m$ or the respective volume constraint if it applies. As an additional criterion, cells that have neighbouring cells with a volume less than 0.1 of their own volume are refined to avoid strong resolution gradients which would degrade the accuracy of the simulation.
We verified that all presented results are unchanged when using $m_0 = 7.5\times 10^5\,\text{M}_\odot$, implying that the results do not depend on the resolution of the background.

We note that in the following `resolution' refers to an additional resolution parameter for the jet material, which has proven to be necessary in these kinds of studies \citep{bourne17}, in particular for low-density jets. This criterion differs for the different models and is not present for the IllustrisTNG kinetic feedback.

\subsection{Equations and models}

We solve the equations of ideal hydrodynamics under the influence of a static external gravitational potential, additionally, radiative cooling from primordial elements and metal lines is included, assuming a constant metallicity of 0.3 times the solar value.
To balance the cooling flow, we use three different feedback injection methods for AGN feedback and assess the uncertainties in modeling and its consequences for galaxy cluster modelling.
First, the jet model presented in \citet[][]{weinberger17b} with some modifications as discussed in the next subsection.
Second, the jet model recently introduced by \citet[][]{su21}, and third, the kinetic AGN wind model used in the IllustrisTNG cosmological simulations \citep{weinberger17}.
While most of the simulations are run with the finite-volume moving mesh code Arepo \citep{springel10}, the runs employing the \citet[][]{su21} model are using the meshless finite mass technique in Gizmo \citep{hopkins15}, thus enabling us to study the effects of differences in hydrodynamics solver (and radiative cooling implementation).

The simulations performed with Arepo rely on the metal cooling and star formation modeling described in \citet{vogelsberger13} with updated parameters from \citet{pillepich18}, consistent with the IllustrisTNG simulations. The simulations performed with Gizmo are using the \citet{hopkins18} model, consistent with the FIRE-2 model.

\subsubsection{RW jet model}

\begin{figure}
    \centering
    \includegraphics{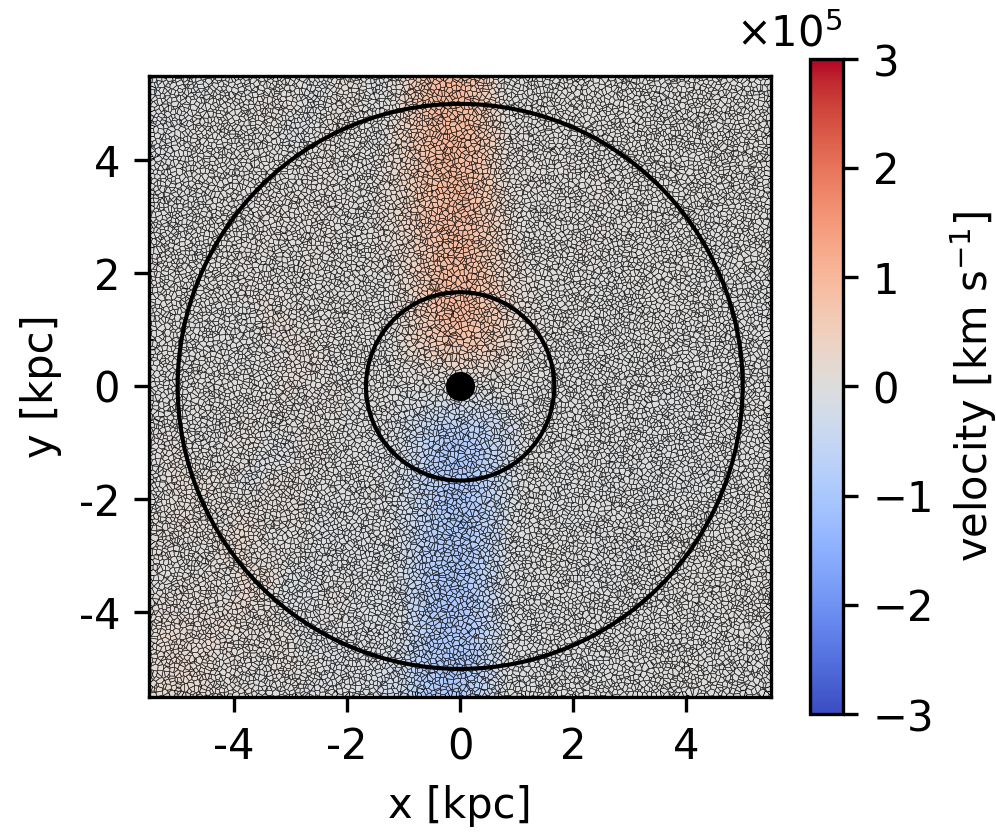}
    \caption{Schematic of the black hole surroundings, including jet (central sphere) and accretion estimate (outer spherical shell) regions. The color-scale indicates the velocity of cells in the jet propagation direction. The schematic shows the highest resolution, \simname{rw jet a 5} simulation in which all regions are well-sampled.}
    \label{fig:jet_schematic}
\end{figure}

For the self-regulated simulations in this work, the algorithm for the jet injection needs to be adjusted from its original implementation in \citet{weinberger17b}.
Originally, the basic idea was to set up a `small scale jet' evolved up to the resolution limit with a clear, separated jet region with density $\rho_\text{jet}$, and the remaining mass distributed into a surrounding `buffer region', taking into account adiabatic compression terms, ensuring exact gas mass conservation in the system, and adding the remaining energy as kinetic energy in the jet.
This approach works well for non-radiative hydrodynamics, but adding mass into an arbitrarily sized buffer region in simulations including radiative cooling leads to spurious effects.
This led to a redesign of aspects of the  algorithm to better fit the needs of self-regulated and ultimately cosmological simulations.

Around the black hole, a radius is defined such that the kernel weighted number of gas cells in this region reaches a predefined value (this is identical to the procedure used e.g. in \citealt{weinberger17}).
Unlike in previous work, this sphere is then separated into an inner sphere with one third of this radius ($r\leq1.65$~kpc), and an outer spherical shell ($1.65$~kpc$<r<5$~kpc).
The inner sphere is used to set up the jet and the outer spherical shell to estimate the surrounding properties; i.e. to calculate the ambient pressure and properties used in the accretion rate estimate. Fig.~\ref{fig:jet_schematic} shows the mesh and velocity field around the black hole and the location of the jet injection region (inner sphere) and outer spherical shell are indicated by circles.

In the inner sphere, we set up gas with density $\rho_\text{jet} = 10^{-28}$~g~cm$^{-3}$ (model \simname{a}), $\rho_\text{jet} = 10^{-27}$~g~cm$^{-3}$ (model \simname{b}), $\rho_\text{jet} = 10^{-26}$~g~cm$^{-3}$ (model \simname{c}), $\rho_\text{jet} = 10^{-25}$~g~cm$^{-3}$ (model \simname{d}). Given the central density of the initial conditions of around $\left<\rho\right> \approx 10^{-25}$~g~cm$^{-3}$, the density contrast $\eta = \rho_\text{jet}\, \left<\rho\right>^{-1}$ ranges from $10^{-3}$ to unity. In the following, we will refer to jets with low $\eta$ as light or low-density and jets approaching $\eta=1$ as dense or heavy.
Mass is removed from (or sometimes, though more rarely, added to) the region to achieve this.
This net removal of mass $\Delta m$ is logged and used in the accretion routine if needed, and $\Delta E$, the thermal energy corresponding to a specific energy $\left<u\right>$ of the surrounding gas is removed from the system, i.e.
\begin{align}
 &\Delta m = \sum\limits_{i} (\rho_i-\rho_\text{jet}) V_i \\
 &\Delta E_\text{mass} = \Delta m \left<u\right>
\end{align}
where the sum is over all cells $i$ in the inner sphere, $V_i$ denotes the volume of cell $i$ and $\left<u\right>$ is a kernel-averaged estimate of the specific thermal energy of the outer spherical shell.

The specific internal energy of the gas in the inner sphere (i.e. the jet region) $u_{\text{jet},i}$ is increased for the gas to be in equilibrium with the pressure in the outer spherical shell $\left<p\right>$.
To ensure no violation of the second law of thermodynamics, and to avoid numerical instabilities, we only allow for increases in specific thermal energy as well as in thermal energy on a per cell basis; i.e. cells that are already over-pressured are kept this way:
\begin{align}
    u_{\text{jet},i} = \max\left(u_i, \left<p\right> \left((\gamma-1) \rho_\text{jet}\right)^{-1}\right), \label{eq:ujet}
\end{align}
where $\gamma$ denotes the adiabatic index of the fluid.
The energy required to establish pressure equilibrium
\begin{align}
    \Delta E_{\text{therm}, i} = \rho_\text{jet} V_i \left(u_{\text{jet}, i}-u_i\right)
\end{align}
is subtracted from the available energy $E_\text{tot}$ and the remaining energy (the dominant part) is used for momentum kicks in bipolar directions,
\begin{align}
    E_\text{jet, kin} = E_\text{tot} - \sum\limits_{i} \Delta E_{\text{mass}, i} - \sum\limits_{i} \Delta E_{\text{therm}, i}.
\end{align}
Note that the momentum is directed in strictly bipolar directions without an opening angle, with the exception of \simname{rw wind}, for which a $60^{\circ}$ opening angle (i.e., velocity kick direction up to $60^{\circ}$ off the jet axis) is applied. The typical velocities at the injection scale can be estimated by
\begin{align}
    \dot{E} \sim \frac{1}{2} \rho_\text{jet} v_\text{jet}^3 A, \label{eq:jet_edot}
\end{align}
where $\dot{E}=10^{45}$~erg~s$^{-1}$ is the jet kinetic luminosity and $A=(1.65\text{ kpc})^2 \pi = 8.6\text{ kpc}^2$ the cross-section of the injection region. The typical velocities are thus $6\times 10^4$~km~s$^{-1}$ (0.2 times the speed of light) for model~\simname{a} ($\rho_\text{jet} = 10^{-28}$~g~cm$^{-3}$) to $6\times10^3$~km~s$^{-1}$ (0.02 times the speed of light) for model~\simname{d} ($\rho_\text{jet} = 10^{-25}$~g~cm$^{-3}$). As shown in Figure~\ref{fig:jet_schematic}, the achieved velocities can be slightly higher due to a weighted injection of kinetic energy which awards on jet-axis cells a larger than average velocity kick and avoids shear discontinuities at injection.

The kinetic luminosity is chosen to match the cooling losses over the simulation time. This approach is motivated by the cumulative energy measured by X-ray cavities roughly balancing cooling losses \citep[e.g.][]{birzan04, olivares22}. Comparing to radio properties of jets \citep[see, e.g.][for a review]{hardcastle20}, this kinetic luminosity lies about an order of magnitude above the FRI/FRII divide in the \citet{kaiser97} model. However, since the jet is propagating in a galaxy cluster, we nonetheless expect the resulting morphology to be substantially affected by its dense environment \citep{owen97, mingo19}. We would thus expect the resulting radio jets to represent the luminous end of the FRI population in dense environments, which are mildly relativistic on kpc scales and decelerate smoothly with substantial entrainment from surrounding gas \citep{bicknell94, laing14} and reach trans-sonic velocities on or slightly beyond kpc scales \citep{bicknell95}.

Note that while some of the jets in our simulations are reaching mildly relativistic speeds, we solve the equations of non-relativistic hydrodynamics in our simulations since we mostly focus on the long-term evolution of the system. The precise morphology and properties of the active jet, however, could be affected by these omitted effects \citep{english16, perucho19, yates22}. For the long-term evolution, however, the ratio of energy to momentum flux carried is crucial, which in the non-relativistic case is proportional to the velocity. All the jets are internally supersonic on kpc scales, with Mach numbers between 1.5 and 4, as detailed in Table~\ref{tab:sims} (see e.g. \citealt{hardcastle14, guo16} on the impact of internal Mach number on lobe properties), which is, for the high resolution runs, slightly higher than the intended transsonic regime and a consequence of avoiding relativistic specific internal energies. We leave exploration of these effects to future work.

The accretion rate is estimated from the outer spherical shell, with the properties such as density and sound speed estimated in a kernel-weighted fashion. From these quantities, the accretion rate $\dot{m}_\text{acc}$ is then calculated using the Bondi accretion rate,
\begin{align}
\label{eq:Bondi}
    \dot{m}_\text{acc} = 4 \pi G^2 m^2_\text{BH} \left<\rho\right> \left<c_\rmn{s}\right>^{-3}
\end{align}
where $G$ is the gravitational constant, $m_\text{BH}$ the black hole mass and $\left<\rho\right>$ and $\left<c_\rmn{s}\right>$ are kernel averaged surrounding density and sound speed, respectively. While the precise functional form of accretion might in practice differ substantially from the employed Bondi formula \citep{gaspari17}, studies with similar setups have found no strong dependence of the self-regulation on the accretion formula as long as it triggers rapid accretion in the presence of cold, dense gas \citep{meece17, ehlert22}.

We note that for the first set of runs we keep the radius of the jet injection routine constant ($1.65$~kpc for the jet region) instead of estimating it from the number of surrounding gas cells. This is done to produce resolution studies that are easier to interpret, in which all effects are attributable to the hydrodynamic resolution of the jet, instead of resolution effects due to the feedback model. We also performed these simulations with a fixed number of cells in the black hole surroundings and hence a resolution dependent physical radius of the injection region of the jet (not shown here) and found no qualitative differences.

Cells that include gas originating from the jet with a mass fraction exceeding $10^{-3}$ are additionally refined to a target volume $V_\text{target}$, thus the mass for these cells is
\begin{align}
    m_i = \rho_i V_\text{target}, \label{eq:jetres}
\end{align}
again refined and derefined accordingly if the actual mass differs by more than a factor of 2. Note that we define the resolution in Table~\ref{tab:sims} as $\rho_\text{jet} V_\text{target}$ to be able to consistently use mass resolution.

\subsubsection{KYS jet model}
\label{subsec:kys_jet_model}

As an independent model comparison, we use the jet launching technique presented in \citet{su21}, utilizing a particle spawning method (see also \citealt{2020MNRAS.497.5292T} and \citealt{wellons2022}).
This method is implemented into the Gizmo code, which solves the Euler equations using a meshless finite mass scheme.
A jet is launched from the center by creating new mass elements at a pre-defined mass flux and attributing a mass, velocity $v_\text{jet}$, temperature $T$ to each resolution element.
To match this model to the previous one, we calculate the temperature using $\rho_\text{jet}$ and external pressure in the initial conditions and the velocity $v_\text{jet}$ using
\begin{align}
    \dot{E} = \frac{1}{2} \rho_\text{jet} v_\text{jet}^3 A + u_\text{jet} \rho_\text{jet} v_\text{jet} A.
\end{align}
To get the mass flux $\dot{M}$ we use
\begin{align}
    \dot{M} = \rho_\text{jet} v_\text{jet} A.
\end{align}

For the model comparison, we use two different jet parameters
\begin{align}
    &\rho_\text{jet} = 10^{-27}\,\text{g~cm}^{-3}; \,\dot{E}=10^{45}\text{ erg s}^{-1} \text{ matched to } \\
    &\qquad v_\text{jet} = 2.6 \times10^4\,\text{km}\,\text{s}^{-1};  \, \dot{M} = 3.5 \,\text{M}_\odot\, \text{yr}^{-1}; \, T = 4.8\times 10^{9}\,\text{K}, \nonumber
\end{align}
for model~\simname{b} and
\begin{align}
    &\rho_\text{jet} = 10^{-26}\,\text{g~cm}^{-3}; \, \dot{E}=10^{45}\text{ erg s}^{-1} \text{ matched to } \\
    &\qquad v_\text{jet} = 1.3 \times10^4\,\text{km}\,\text{s}^{-1};  \, \dot{M} = 17 \,\text{M}_\odot\,\text{yr}^{-1}; \, T = 4.8\times 10^{8}\,\text{K}, \nonumber
\end{align}
for model~\simname{c}. The internal Mach number in the jet is between $3$ and $4.5$ at the injection scale, as detailed in Table~\ref{tab:sims}. The mass of each spawned particle $i$ is $m_i= 4.5\times10^2$~M$_\odot$ and $m_i= 4.5\times10^3$~M$_\odot$ for the two runs, respectively.\footnote{Note that the mass per jet resolution element in the simulations run with Gizmo is higher than for the Arepo simulations in order to ensure both are converged.}

\subsubsection{TNG kinetic winds}
As a comparison with established models that were previously used in cosmological simulations, we run simulations using the kinetic wind model of \citet{weinberger17}.
Unlike in the cosmological context, we enforce the model to be always in kinetic wind and never in thermal mode.
This is done because the kinetic wind mode has been shown to be the dominant one in quenching of massive galaxies and keeping galaxies quiescent \citep{weinberger18}. The kinetic wind mode is thus also responsible for reducing cooling flows.

The injection region for the wind is a sphere around the black hole with radius $h$
\begin{align}
  n_\mathrm{ngb} = \sum\limits_i \frac{4\,\pi}{3} \frac{h^3 m_i}{m_0} w(r_i)   \label{eq:tngnngb}
\end{align}
where the number of neighbouring gas cells $n_\mathrm{ngb} = 64$ is a free parameter, $m_i$ the mass of cell $i$, $m_0$ the target gas mass and $w(r_i)$ a (SPH-like) cubic spline kernel with dimensions of an inverse volume and softening length $h$ and $r_i$ the distance of the cell to the black hole. In the case of the fixed kinetic luminosity runs, the available energy $\Delta E$ simply accumulates, in the case of the self-regulated simulation,
\begin{align}
    \Delta E = \int \epsilon_\text{f,kin} \dot{m}_\text{acc} c^2 dt  \label{eq:tngdeltae}
\end{align}
where $\epsilon_\text{f,kin} = 0.1$, $\dot{m}_\text{acc}$ is the accretion rate determined by the Bondi-accretion rate formula in Equation~\eqref{eq:Bondi} and $c$ denotes the speed of light. The injection of the wind happens in a pulsed fashion, with a pulse being injected if the accumulated energy $\Delta E$ is equal or exceeds
\begin{align}
    E_\mathrm{inj, min} = f_\mathrm{re} \, \frac{1}{2} \sigma_\mathrm{DM}^2 m_\mathrm{enc}  \label{eq:tngemin}
\end{align}
where in this simulation we use a fixed $\sigma_\mathrm{DM} = 3.3\times10^2$~km~s$^{-1}$ since the presented simulations do not contain live dark matter particles to be measured on the fly. $m_\mathrm{enc}$ is the enclosed gas mass within the injection sphere and $f_\mathrm{re}=20$ a free parameter controlling the frequency of pulses. Once the available energy $\Delta E$ exceeds $E_\mathrm{inj, min}$, a momentum kick
\begin{align}
    \Delta \mathbfit{p}_i = m_i \sqrt{\frac{2 \, \Delta E \, w(r_i)}{\left<\rho\right>}}\hat{\mathbfit{n}}  \label{eq:tngdp}
\end{align}
is applied to all cells $i$ within the injection sphere. Note that there is a single injection direction $\hat{\mathbfit{n}}$ with no opening angle. However, the injection direction is chosen randomly from a unit sphere after each pulse, resulting in an spherical injection on average.

\begin{table}
    \centering
    \begin{tabular}{c|c}
        \hline
        model parameters &  \\
        \hline
        $n_\mathrm{ngb}$ & $64$ \\
        $f_\mathrm{re}$ & $20$ \\
        $\sigma_\mathrm{DM}$ & $3.3\times 10^2$~km~s$^{-1}$ \\
        $\epsilon_\text{f,kin}$ & $0.1$ \\
        \hline
        wind properties (derived) & \\
        \hline
        $v_\mathrm{wind}$ & $1.5\times 10^3$ \\
        $\rho_\mathrm{wind}$ & $10^{-25}$~g~cm$^{-3}$ \\
        $\dot{m}_\mathrm{wind}$ & $7$~M$_\odot$~yr$^{-1}$ \\
        \hline
    \end{tabular}
    \caption{Model and wind parameters of the TNG kinetic wind model. The model parameters are defined in equations (\ref{eq:tngnngb})-(\ref{eq:tngemin}), the wind velocity $v_\mathrm{wind}$, density $\rho_\mathrm{wind}$ and mass flux $\dot{m}_\mathrm{wind}$ are derived using a single injection event and deriving characteristic quantities (using the typical injection sphere radius of $h=1$~kpc).}
    \label{tab:tngparam}
\end{table}

These parameters imply that the small scale wind of an individual injection event reaches velocities of $1.5\times10^3$~km~s$^{-1}$. The model uses the ambient density and only adds momentum kicks without injecting mass, making a definition of a mass flux somewhat ambiguous. Using the initial surrounding density of $10^{-25}$~g~cm$^{-3}$ and the injection region radius of about $1$~kpc, this results in a mass flux of $7$~M$_\odot$~yr$^{-1}$.
In summary, the parameters are close, but not identical to IllustrisTNG.

\subsection{Summary of simulations}
\begin{table*}
    \centering
    \begin{tabular}{c|c|c|c|c|c|c}
        \hline
         \textbf{single outburst} \\
         \hline
         Name & Model & Time & Jet density $\rho_\mathrm{jet}$ & Relative density $\eta$ & Jet Mach number $\mathcal{M}$ & Resolution\\
         \hline
         rw jet a 1 & rw17 & 250 Myr & $10^{-28}$~g~cm$^{-3}$ & $10^{-3}$ & $1.5$ & $2.5 \times 10^3$~M$_\odot$\\
         rw jet a 2 & rw17 & 250 Myr & $10^{-28}$~g~cm$^{-3}$ & $10^{-3}$ & $2.0$ & $3.1 \times 10^2$~M$_\odot$\\
         rw jet a 3 & rw17 & 250 Myr & $10^{-28}$~g~cm$^{-3}$ & $10^{-3}$ & $2.5$  & $39$~M$_\odot$\\
         rw jet a 4 & rw17 & 250 Myr & $10^{-28}$~g~cm$^{-3}$ & $10^{-3}$ & $3.0$ & $4.9$~M$_\odot$\\
         rw jet a 5 & rw17 & 25 Myr & $10^{-28}$~g~cm$^{-3}$ & $10^{-3}$ & $3.0$ & $0.61$~M$_\odot$\\
         rw jet b 2 & rw17 & 250 Myr & $10^{-27}$~g~cm$^{-3}$ & $10^{-2}$ & $2.5$ & $9.4 \times 10^3$~M$_\odot$\\
         rw jet c 2 & rw17 & 250 Myr & $10^{-26}$~g~cm$^{-3}$ & $10^{-1}$ & $3.0$ & $9.4 \times 10^4$~M$_\odot$\\
         rw jet d 2 & rw17 & 250 Myr & $10^{-25}$~g~cm$^{-3}$ & $1$ & $4.0$ & $3.1 \times 10^5$~M$_\odot$\\
         rw wind d 2 & rw17 & 250 Myr & $10^{-25}$~g~cm$^{-3}$ & $1$ & $4.0$ & $3.1 \times 10^5$~M$_\odot$\\
         tng & rw17a & 250 Myr & ambient & $1$ & - & $9.4 \times 10^4$~M$_\odot$ \\
         rw no jet & - & 250 Myr & - & - & - & $9.4 \times 10^4$~M$_\odot$ \\
         \hline
         kys jet b & su21 & 250 Myr & $10^{-27}$~g~cm$^{-3}$ & $10^{-2}$ & $3.0$ & $4.5 \times 10^2$~M$_\odot$\\
         kys jet c & su21 & 250 Myr & $10^{-26}$~g~cm$^{-3}$ & $10^{-1}$ & $4.5$ & $4.5 \times 10^3$~M$_\odot$\\
         kys no jet & - & 250 Myr & - & - & - & $9.4 \times 10^4$~M$_\odot$ \\
         \hline
         \textbf{self-regulated} \\
         \hline
         jet & rw17 & 2000 Myr & $10^{-28}$~g~cm$^{-3}$ & $10^{-3}$ & variable & $3.1 \times 10^2$~M$_\odot$\\
         tng & rw17a & 2000 Myr & ambient & $1$ & - & $9.4 \times 10^4$~M$_\odot$ \\
         no AGN & - & 2000 Myr & - & - & - & $9.4 \times 10^4$~M$_\odot$ \\
         \hline
    \end{tabular}
    \caption{List of simulations and their parameter variations.  From left to right, we show the name, the model (rw17 refers to \citealt{weinberger17b}, rw 17a to \citealt{weinberger17}, su21 to \citealt{su21}) the simulated time, the jet density $\rho_\mathrm{jet}$, which is an input parameter, the approximate relative density ratio relative to the central density in the initial conditions $\eta$ and the measured jet Mach number $\mathcal{M}$ on kpc scales after $6.25$~Myr and finally the mass resolution in the jet region. All single outburst calculations have a jet kinetic luminosity $\dot{E}=10^{45}$~erg~s$^{-1}$ for a time of $25$~Myr.  The simulations starting with kys are performed with the Gizmo code, all other runs with Arepo.}
    \label{tab:sims}
\end{table*}

The main challenge in this study is the dynamic range in time. High spatial resolution in the jet requires very short timesteps, while covering several central cooling times requires an overall simulation time of order several Gyr. Even with state of the art computing capabilities and numerical optimizations, this is challenging and would prohibit model variations.

To overcome these difficulties, we use two different types of simulations \textit{starting from the same initial conditions}. First, we run fixed jet power simulations with $\dot{E} = 10^{45}$~erg~s$^{-1}$, beginning from a single outburst lasting for $25$~Myr; i.e. a total energy injection of $8\times 10^{59}$~erg, and an overall simulation time up to $250$~Myr.\footnote{The highest resolution jet stops at $25$~Myr due to computational limitations.} Note that the cooling luminosity in the halo is approximately $10^{44}$~erg~s$^{-1}$. These simulations will be analysed in Sections~\ref{sec:jet_propagation} and \ref{sec:effect_on_halo}. Second, we run a few self-regulated simulations in which the feedback energy is calculated using the accretion rate. These runs are evolved for $2$~Gyr and establish an equilibrium between cooling flow and star formation. For the latter simulation, we compare two models, the IllustrisTNG kinetic wind (\simname{tng}) and the \simname{rw jet a 2} model (\simname{jet}). These simulations will be shown in Section~\ref{sec:self_regulated}.
A summary of the simulations can be found in Table~\ref{tab:sims}. We restrict the resolution study to model \simname{rw jet a} since this lowest-density jet tends to be most sensitive to changes of resolution.

Note that all these simulations use non-relativistic ideal hydrodynamics for simplicity. Magnetic fields are not included, yet for all 3 models studies highlighting their effect exist. \citet{weinberger17b} showed that while a weak magnetization increases stability of the lobes, even hydrodynamical lobes of the jets presented in this work are stable over the relevant timescales. \citet{su21} found magnetic field to play a minor role for feedback unless the injected magnetic fluxes are $>10^{44}$~erg~s$^{-1}$ in a setup similar to the one presented in this work. Finally, \citet{ehlert22} studied self-regulated cool core systems and found magnetic fields to dominate the dynamics of cold gas, while they connect the cold to the hot ICM phase, thereby enabling efficient (angular) momentum transport from one phase to the other \citep{wang2021}. This effect seems to be important for shaping the thermodynamics and kinematics of the cold phase on timescales relevant for long-time self-regulation of the ICM \citep{ehlert22}, which goes beyond the focus of this study. Relativistic effects to the equation of state of the jet material or its cooling functions are not included. The jet material is just modeled as very hot, relatively low density and thus radiatively inefficient plasma. Overall, we do not expect the results presented in the following to be substantially affected by these simplifications.
\section{Jet propagation}
\label{sec:jet_propagation}

\begin{figure}
    \centering
    \includegraphics{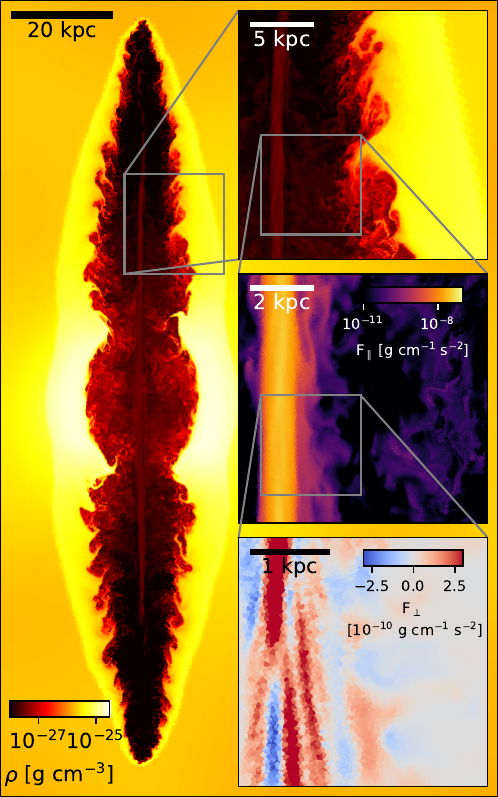}
    \caption{Density slice (main panel and upper inlay) of \simname{rw jet a 5} after $25$~Myr. The central inlay shows the momentum flux density in the jet direction $F_\parallel$, the bottom panel shows the momentum flux perpendicular to the jet direction $F_\perp$. Its change of sign indicates body-modes in the jet flow. The shape of the resulting cavity differs from observed x-ray cavities likely due to the absence of density fluctuations in its propagation direction as well as its constant luminosity.}
    \label{fig:jet_slice}
\end{figure}

We begin our analysis of AGN feedback with the propagation of a low-density jet of fixed luminosity.
Jet propagation is a key aspect for feedback since the spatial deposition of feedback energy will be decisive for the jet's ability to impact cooling flows and not just heat the already hot, low density outer ICM.
We therefore hypothesize that jet propagation needs to be converged in order to obtain converged feedback effects from the jets.
The general problem of jet propagation is already covered in \citet{weinberger17b}, and the following section expands on this showing jets with higher resolution and in simulations with radiative cooling.

Figure~\ref{fig:jet_slice} shows a slice though the highest resolution jet simulation (\simname{rw jet a 5}) at $25$~Myr; i.e. when the jet shuts off.
The main panel and the first zoom-in show the density, highlighting the cocoon as well as the bow shock.
The other two inlay panels show the component of the momentum flux density in jet direction transported in jet direction ($F_\parallel$, middle) and perpendicular to the jet direction ($F_\perp$, bottom), which highlight the jet propagation in the inflated cocoon.
While the formation of the low Mach number bow shock \citep{ehlert18} is comparably simple to resolve and generic to outflows \citep{fauchergigere12, king15, costa20}, the contact discontinuity, i.e. the transition between shocked, dense material and shocked, under-dense jet material forming a cavity, is prone to subsonic Kelvin-Helmholtz instabilities (KHI) and requires higher resolution.
The extent to which these instabilities can grow depends on the ratio of the growth time to the dynamical time of the cavity itself, set by the buoyancy timescale \citep{weinberger17b}. Resolving this mixing process between cavity and shocked ICM represents a significantly more difficult challenge computationally, since the relevant modes of the KHI need to be resolved.

Finally, the middle and lower panels show the jet as a collimated supersonic flow propagating in its cavity.
Maintaining the momentum flux confined in a cylinder with diameter of only a few kpc is only possible if the velocity shear layer is resolved, and if the hydrodynamic flow does not develop instabilities on timescales shorter than the jet propagation time.
In practice, this is the case if the flow is internally supersonic (see discussion in \citealt{Padnos2018} and \citealt{mandelker19}, their Section~2, as well as \citealt{Berlok2019a,Berlok2019b} for the magnetised case), and if the jet is resolved linearly by at least of order 10 cells per diameter, making this the most challenging aspect to model accurately in numerical simulations.
If the outward momentum transport is not captured accurately in a numerical simulation, the respective position of the jet will be incorrect.
Inverting this argument, a converged position of the jet head implies an accurate modeling of the momentum flux in the jet\footnote{Within the assumptions of an ideal gas. This certainly falls short to explain real AGN driven jets.}. Note also that the propagation in this idealized setting with an initially spherically symmetric ICM density profile and constant jet luminosity is likely to lead to substantially larger travel distances and consequently more elongated cavities than comparable jets in cool-core galaxy clusters \citep{owen97}.

\begin{figure*}
    \centering
    \includegraphics{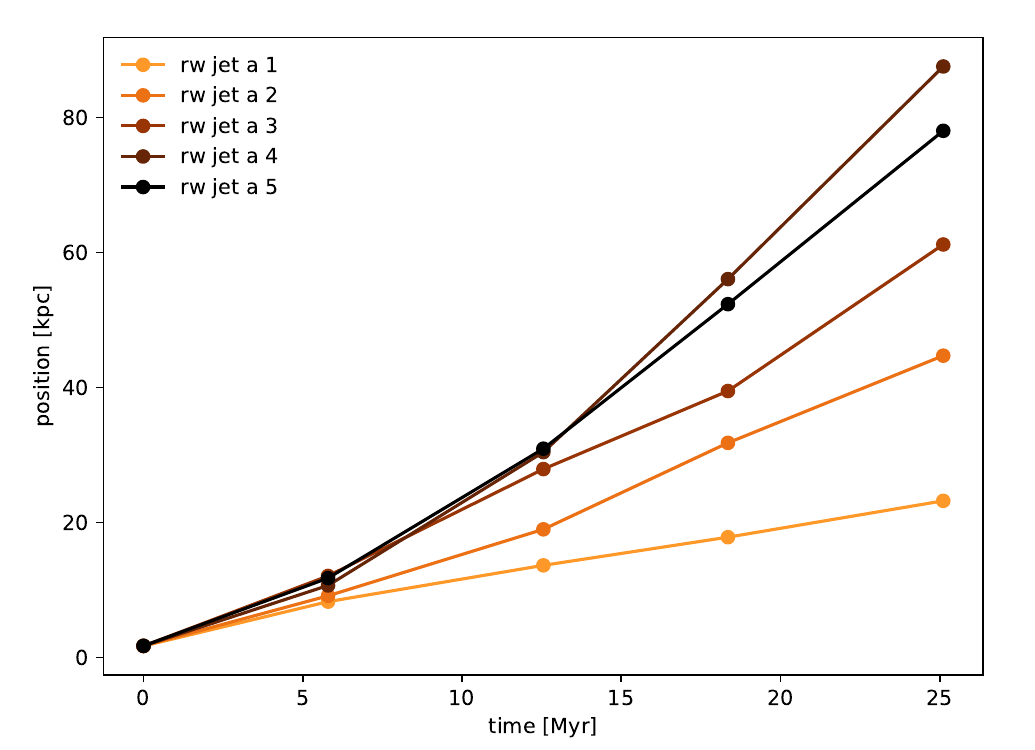}
    \caption{Jet position vs.\ time for different jet and ICM resolutions.
    This shows that it is necessary to resolve the jet to resolution level 4, i.e., with resolution elements smaller than $100$~pc for the jet propagation to be resolved.}
    \label{fig:pos_time}
\end{figure*}

We use the position of the jet at a given time as a measure of convergence of jet propagation, as shown in  Fig.~\ref{fig:pos_time}.
It turns out that this seemingly simple criterion is non-trivial to achieve, even for hydrodynamic jets \citep{weinberger17b, yates18}, with jet distance generally increasing with increased resolution.
Pushing the resolution in the jet further than in previous studies, however, we can see a turnaround in jet distance vs.\ time plot with increasing resolution, in particular between \simname{rw jet a 4} and \simname{rw jet a 5} in Fig.~\ref{fig:pos_time}.
The colors indicate the different jet resolutions. Comparing the slices in Fig.~\ref{fig:jet_slice} (which shows \simname{rw jet a 5}) with the equivalent plot for \simname{rw jet a 4} (Fig.~\ref{fig:jet_slice_jet4}), the most significant difference is the absence of body modes in the transverse momentum flux (bottom inlay panel) in the lower resolution run. We speculate that this momentum transport via transverse body modes of the KHI takes over the slowing down of the jet while it is dominated by numerical viscosity or numerical diffusion at lower resolutions.

\section{Effect on ICM}
\label{sec:effect_on_halo}

\begin{figure}
    \centering
    \includegraphics{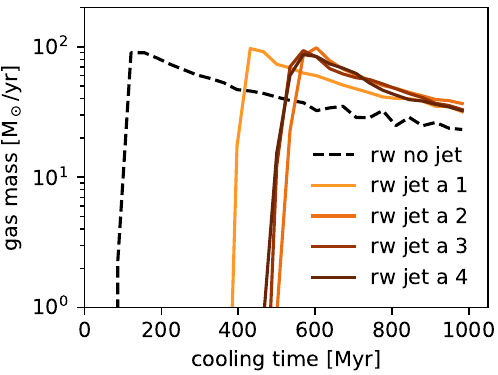}
    \caption{Gas cooling time histogram for different jet resolutions after 50~Myr. The `preventive feedback' effects of jets, i.e. the delay of the developing cooling flow compared to the \textit{no jet} simulation, are converged starting from the resolution of \textit{rw jet a 2}, i.e. earlier compared to jet position.}
    \label{fig:tcool_resolution}
\end{figure}

\begin{figure}
    \centering
    \includegraphics{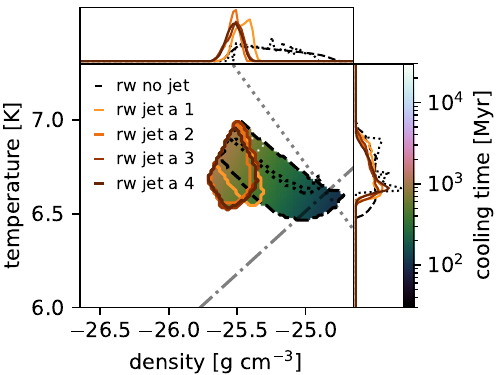}
    \caption{Phase diagram of cooling gas ($t_\text{c} < 1$~Gyr) for runs with different jet resolutions after 50~Myr. The dash-dotted line indicates an adiabat at $2$ keV cm$^2$, the dotted line an isobar at $2\times 10^{-10}$~erg~cm$^{-3}$. The reduction of cooling time is related to the boosting of the gas entropy, indicating an immediate response of the halo gas to the jet, rather than an isochoric temperature increase of the central gas. The black dotted contour shows the initial conditions.}
    \label{fig:phase_diagram_resolution}
\end{figure}

\begin{figure*}
    \centering
    \includegraphics{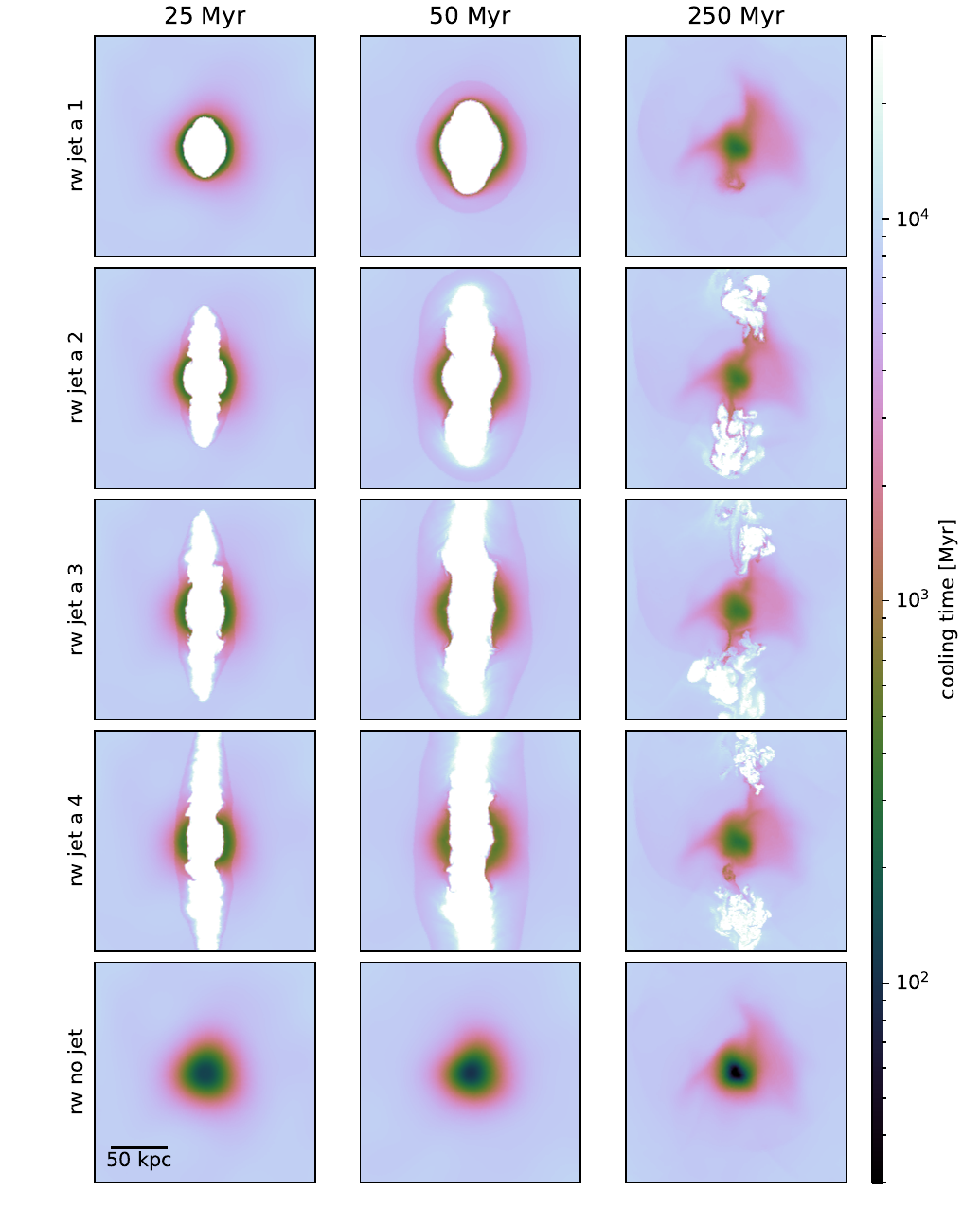}
    \caption{Slices of the cooling time of simulations with different jet resolutions (rows) after $25$, $50$ and $250$~Myr (columns). The maps show a region of $200$~kpc side length. The last row shows a simulation without a jet, for reference. Except for the lowest jet resolution (\simname{rw jet a 1}), the effects of the jet on cooling time are very similar due to similar behaviours in the central region, despite different jet propagation further out.}
    \label{fig:tcool_slices_resolution}
\end{figure*}

\begin{figure*}
    \centering
    \includegraphics{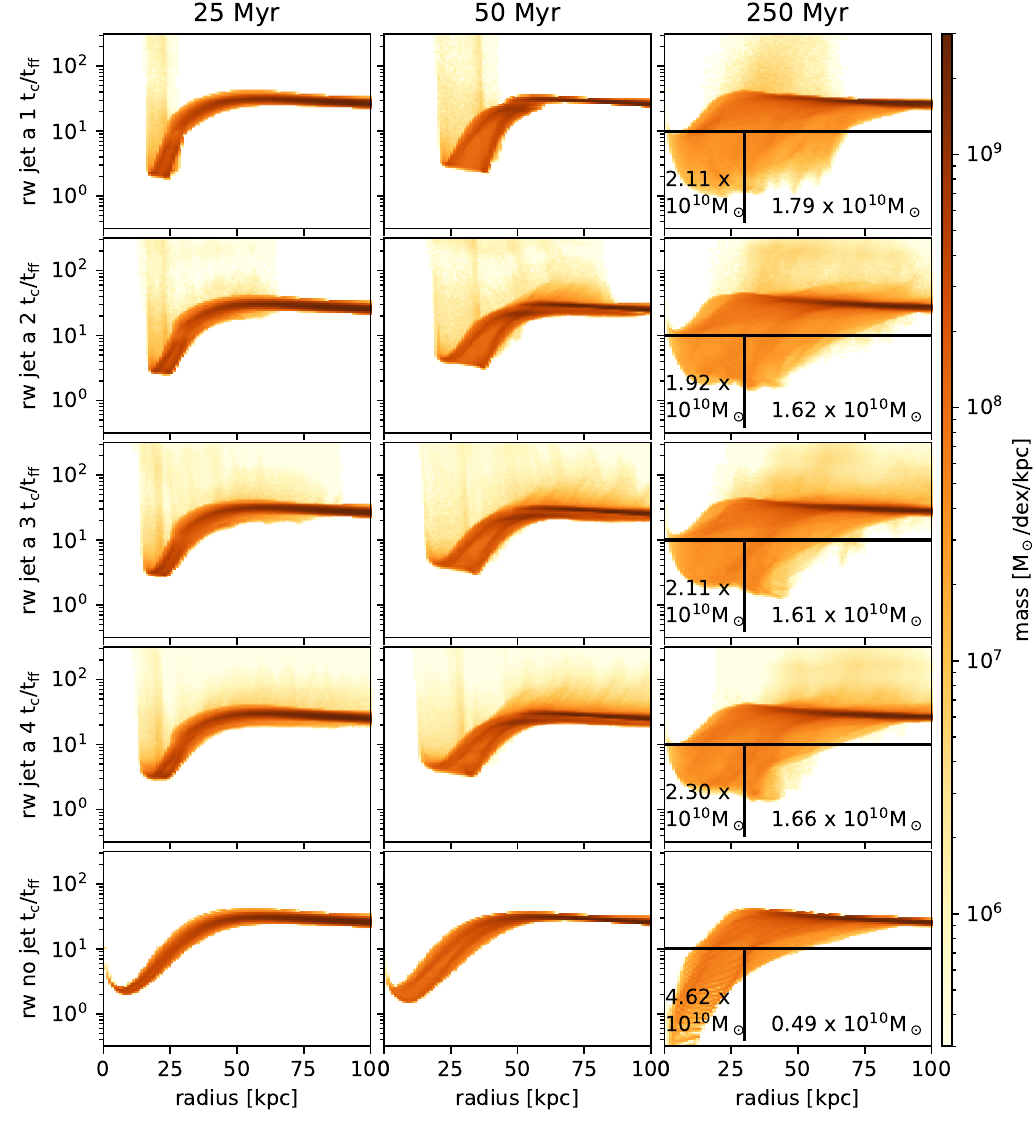}
    \caption{Two-dimensional histogram of cooling over free-fall time as a function of radius for different jet resolutions.}
    \label{fig:tcool_tff_profile_resoltion}
\end{figure*}

\begin{figure}
    \centering
    \includegraphics{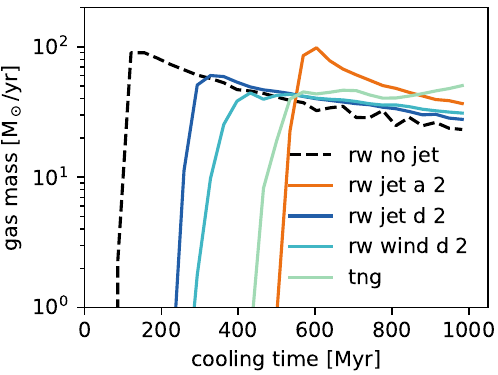}
    \caption{Same as Fig.~\ref{fig:tcool_resolution}, only with different AGN feedback models after 50~Myr. The ability to alter the cooling time PDF, and consequently the ability to moderate cooling-flows is highly model and parameter dependent.  The resolution effects  for the \simname{rw jet d 2} are smaller than for the light jet indicated in Fig.~\ref{fig:tcool_resolution} (not shown).}
    \label{fig:tcool_models}
\end{figure}

\begin{figure}
    \centering
    \includegraphics{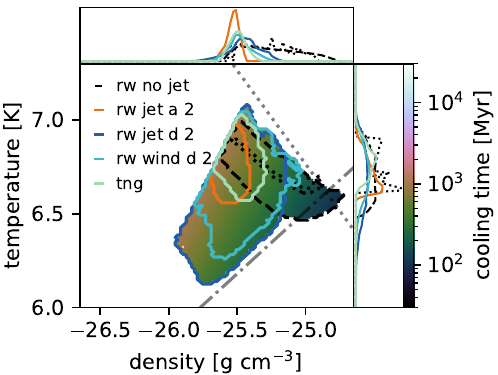}
    \caption{Same as Fig.~\ref{fig:phase_diagram_resolution} but comparing the simulations with varying model parameters.}
    \label{fig:phase_diagram_model}
\end{figure}

\begin{figure*}
    \centering
    \includegraphics{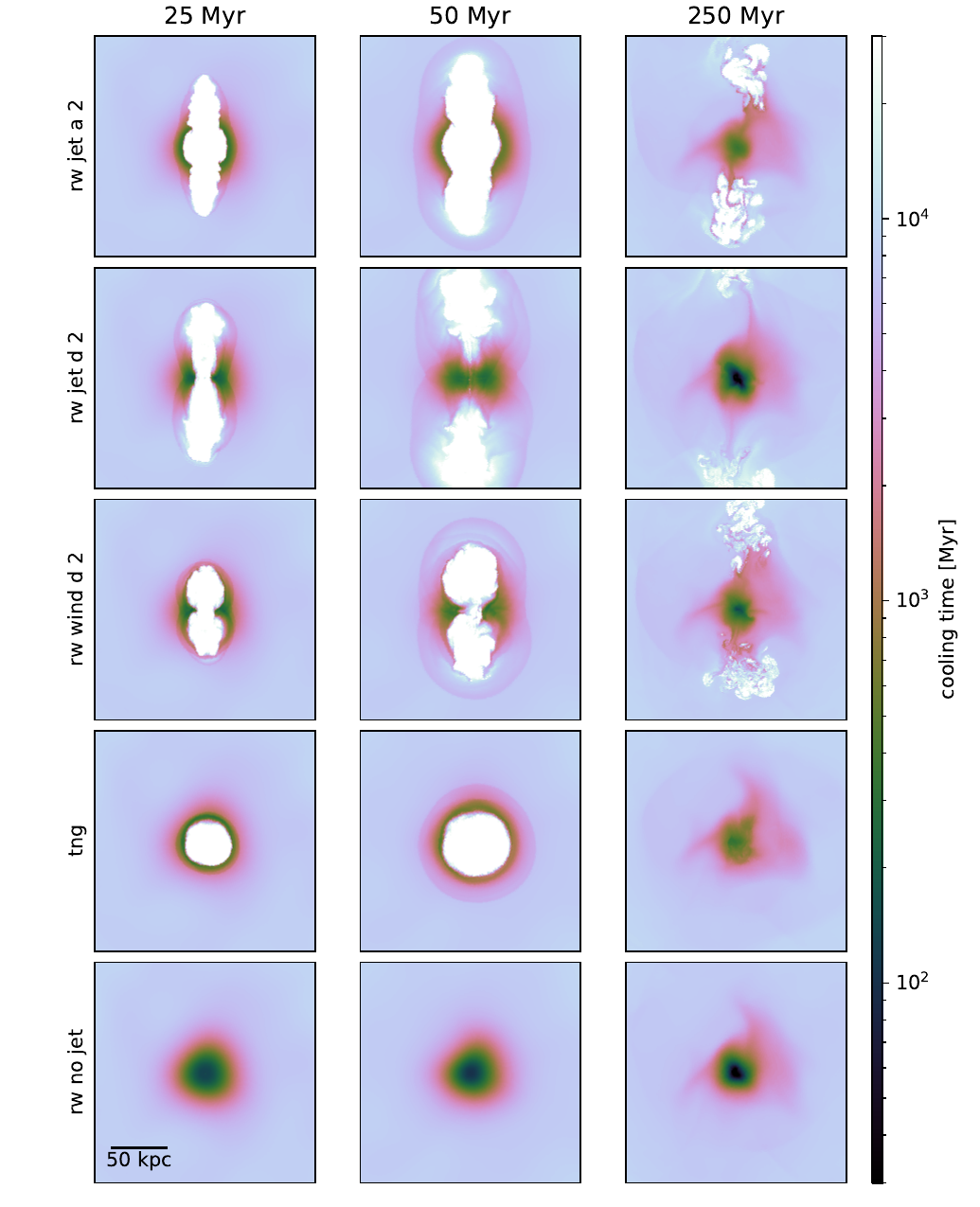}
    \caption{Same as Fig.~\ref{fig:tcool_slices_resolution}, but for simulations with different jet parameters and feedback models. The last row shows the simulation without feedback. The inefficiency of \simname{rw jet d 2} to offset cooling flows seems to originate from its inability to affect the central gas.}
    \label{fig:tcool_slices_model}
\end{figure*}

\begin{figure*}
    \centering
    \includegraphics{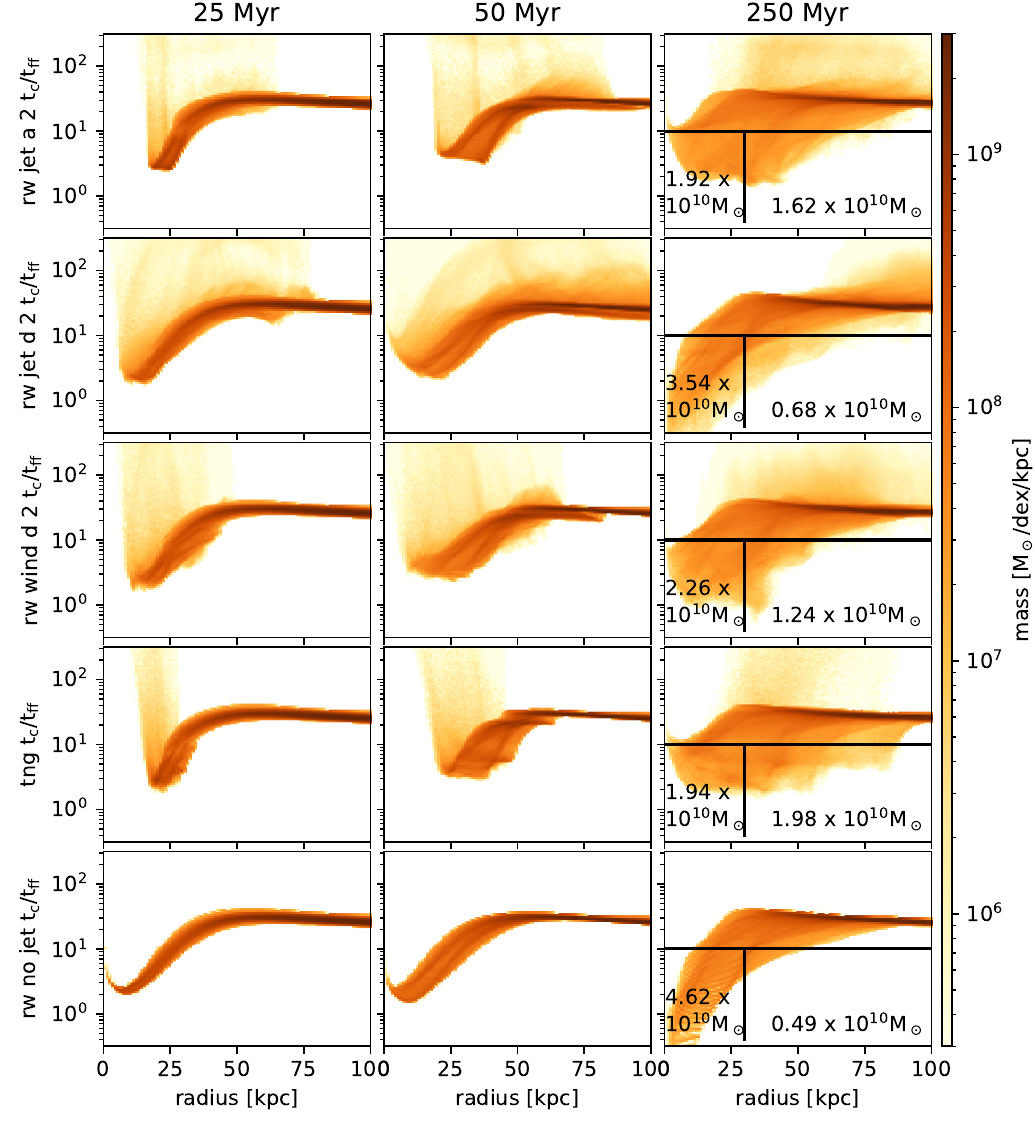}
    \caption{2d histogram of cooling over free-fall time as a function of radius for different AGN driven jet and wind models.}
    \label{fig:tcool_tff_profile_model}
\end{figure*}

\begin{figure}
    \centering
    \includegraphics{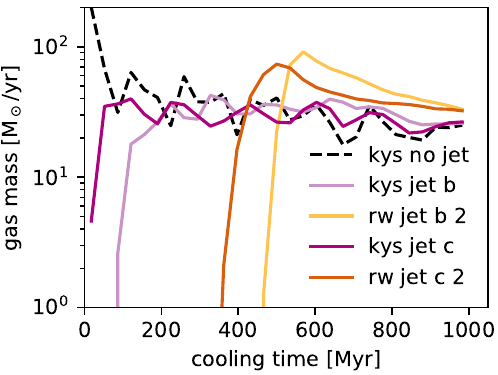}
    \caption{Same as Fig.~\ref{fig:tcool_resolution}, only with different AGN feedback model and run with a different code, cooling and star formation model. Simulations starting with \simname{kys} use the meshless finite mass mode of the Gizmo code, while simulations starting with \simname{rw} employ a finite volume moving mesh solver using Arepo. The ability to alter the cooling time PDF, and consequently the ability to moderate cooling-flows is highly code as well as model parameter dependent.}
    \label{fig:tcool_code}
\end{figure}

\begin{figure}
    \centering
    \includegraphics{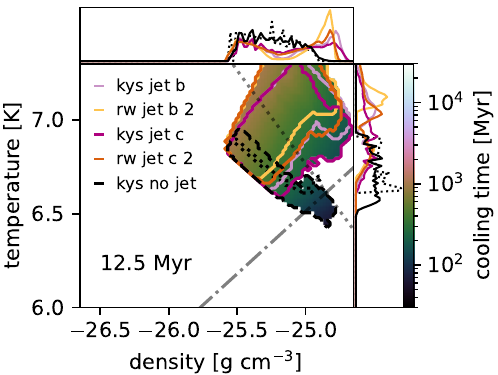}
    \includegraphics{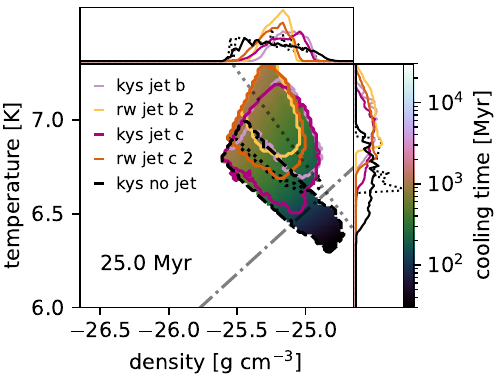}
    \includegraphics{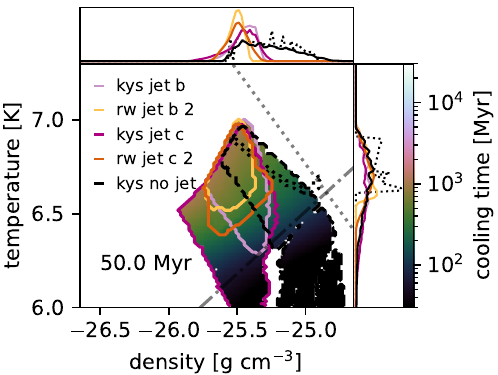}
    \caption{Phase diagram of gas with short cooling time after 12.5~Myr (top), 25~Myr (middle) and 50~Myr (bottom) for simulations with different models and using different simulation codes. Simulations starting with \simname{kys} use the meshless finite mass mode of the Gizmo code, while simulations starting with \simname{rw} employ a finite volume moving mesh solver using Arepo. The increased difference in gas distribution between the different codes at later times indicates the role of the different cooling functions and the non-linearity of the developing cooling flow.}
    \label{fig:phase_diagram_code}
\end{figure}

\begin{figure*}
    \centering
    \includegraphics{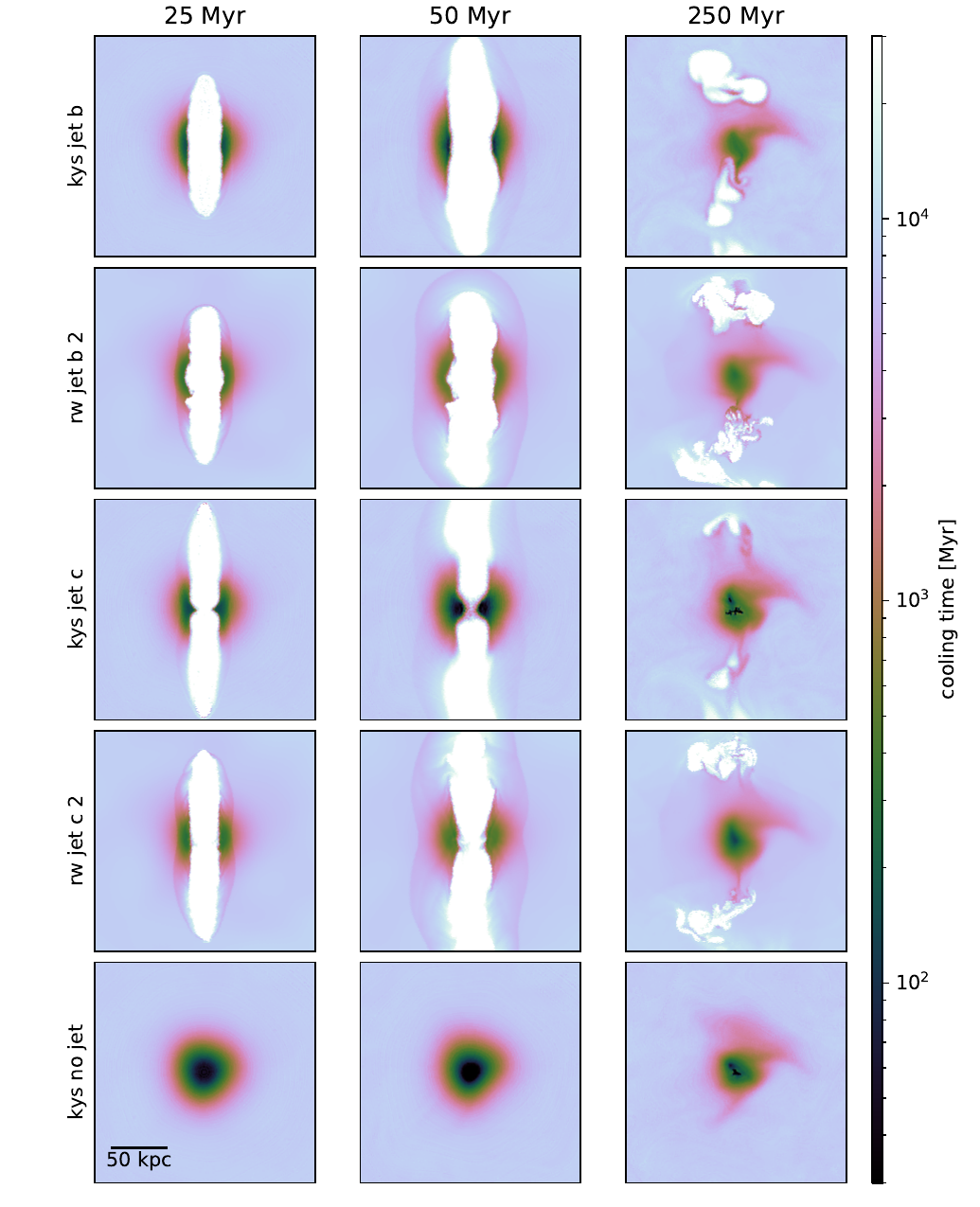}
    \caption{Same as Fig.~\ref{fig:tcool_slices_resolution}, but for simulations with different codes and jet implementations. Simulations starting with \simname{kys} use the meshless finite mass mode of the Gizmo code, while simulations starting with \simname{rw} employ a finite volume moving mesh solver using Arepo. The last row shows the simulations without feedback (Gizmo version).}
    \label{fig:tcool_slices_code}
\end{figure*}

Having established the jet propagation and its behaviour with resolution, we now move on to study the effect of the jet event on the hydrostatic halo.
We are particularly interested in the jet's ability to moderate or delay a developing cooling flow, and its dependence on numerical and modelling aspects.
For this purpose, we use the fixed luminosity jet simulations (as in the previous section) with a jet kinetic luminosity of $10^{45}$~erg~s$^{-1}$ for $25$~Myr. These simulations include radiative cooling which allows us to study the effect of the jet on cooling times.

\subsection{Numerical convergence of AGN jet feedback}
Figure~\ref{fig:tcool_resolution} shows the gas mass-weighted cooling-time distribution function for different jet resolutions and a reference run without a jet (black dashed line).
All runs with jets cause a significant shift of the distribution function towards longer cooling times.
\simname{rw jet a 1} manages to shift it by $300$~Myr, while starting at the resolution of \simname{rw jet a 2}, the cooling time distribution function converges, being shifted by $400$~Myr relative to the reference run. The radiated energy over $400$~Myr (with cooling losses of order $10^{44}$~erg~s$^{-1}$) corresponds to $1.3\times 10^{60}$~erg, which even slightly exceeds the injected energy, $0.8 \times 10^{60}$~erg. This proves that jets with an average power that roughly matches the cooling luminosity are capable of delaying cooling flows.
To determine the physical reason for this delay, we plot the phase diagram of the low cooling time gas ($< 1$Gyr)  in Fig.~\ref{fig:phase_diagram_resolution}. Interestingly, the gas in the presence of jets is not hotter than the gas in the no jet case, but the high density, high pressure end is removed. Given the strong dependence of radiative cooling on density, it is not surprising that this leads to a substantial reduction in cooling luminosity and an absence of gas with short cooling times.
We have thus shown that light hydrodynamical jets are able to efficiently delay developing cooling flows, mostly by removing the dense gas component. While the steep density dependence of the cooling time is the dominant factor for delaying the cooling flow, the removal of dense gas does involve non-adiabatic processes on the lowest entropy gas in the center.
Note that we do not make any statement about the energy coupling mechanisms to the ICM. However, these results suggest that the feedback effects of a jet after $50$~Myr, i.e. $25$~Myr after the jet shut off is not an isochoric heating process, i.e., an increase central gas temperature at fixed density, but already mediated by an increase in the entropy of the gas which reduces further cooling. Substantial increases in pressure, and consequently strong shocks are absent at this point. We will closer examine the time evolution of the gas state at times $<50$~Myr in section \ref{sec:code}.

Examining the resolution dependence, it is particularly noteworthy that the thermodynamic effects on the ICM seem to converge at lower resolution than the jet propagation itself.
This seemingly puzzling result can be easily illustrated with a map of the cooling times in the central cluster regions.
Figure~\ref{fig:tcool_slices_resolution} shows the cooling time at different times (columns) and for different resolutions (rows), with the last row showing an unmediated cooling flow for reference. Note that the middle column corresponds to the time when we analyze the cooling time distribution function in Figs.~\ref{fig:tcool_resolution} and \ref{fig:phase_diagram_resolution}.
The low cooling time gas is all centrally concentrated at distances smaller than the jet travel distance.
Thus, the exact jet traveling distance has no immediate impact on the low cooling time gas.
The notable exception is \simname{rw jet a 1}, where the jet is not able to break out of the inner region of cold gas at all.

What about the effect of resolution to the gas further out?
Figure~\ref{fig:tcool_tff_profile_resoltion} shows a mass-weighted 2d histogram of cooling over free-fall time as a function of radius $r$. The free-fall time is defined as $t_\text{ff} = \sqrt{2\,r\,g^{-1}}$, with $g$ being the radial gradient of the gravitational potential.
The gas above the main ridge is the hot jet cocoon material and the different propagation behaviour between the models can be clearly identified in the left and central columns. Interestingly, however, this does not result in significantly different cooling over free-fall time distributions.
In particular the mass of gas outside $30$~kpc with $t_\text{c}/t_\text{ff} < 10$ is converged to a few percent starting at \simname{rw jet a 2}.
Notably, the amount of cold gas in the outskirts increases due to the presence of a central jet, while the reference run develops a significant cooling flow and a factor of several higher cooling luminosity by $250$~Myr, however, only in the center. This effect is caused by substantial (ICM) gas outflows driven by the jet and the subsequent rising of the lobe \citep[not explicitly shown here, but see e.g.][]{chen19, prasad21}.

\subsection{Sensitivity to modelling of jet feedback}
\label{sec:modelvar}

Having shown numerical convergence, we now examine the dependence on different physical models for jets and AGN driven outflows. For the model variations, we subsequently use resolution level 2 for which convergence has been established.
Figure~\ref{fig:tcool_models} shows the cooling time distribution function for a variety of different models, in particular a jet with higher density (\simname{rw jet d 2}), and an outflow with higher density and wide opening angle (\simname{rw wind d 2}).
For reference, we also compare this to the IllustrisTNG kinetic wind mode (\simname{tng}) which acts via frequent momentum kicks of the central gas in a random direction.

Most notably, the impact of model differences is substantially larger than the resolution variations of \simname{rw jet a} discussed in Fig.~\ref{fig:tcool_resolution}. The resolution effects of the \simname{rw jet d} and \simname{rw wind d} tend to be even smaller. This implies that the underlying modelling assumptions in the feedback model are the biggest source of uncertainty in this setup (note that this setup uses a fixed feedback power and duration and does not contain any complex multi-phase gas).
Looking at the individual simulations, there are a number of noteworthy trends. In particular, heavy jets have substantially weaker effects than light, low-density jets.
This effect can be somewhat overcome by choosing an opening angle (\simname{rw wind d 2}), yet the basic problem persists.
The \simname{tng} run exhibits a similar removal of low cooling time gas as the light jet model.
Figure~\ref{fig:phase_diagram_model} shows the phase diagram of the gas with low cooling times for the different models. The ability to delay the cooling flow seems to correlate more with the absence of dense than with cold gas. \simname{rw jet d 2} and \simname{rw wind d 2} have a reduced ability to mediate the cooling flow. The \simname{tng} model has a similar effect as \simname{rw jet a 2} in this setup.

To understand this outcome better, it is again useful to look at maps of the cooling time for the different runs, which are shown in Fig.~\ref{fig:tcool_slices_model}. The similarity in cooling times of the \simname{rw jet a 2} and \simname{tng} models after $50$~Myr (first and fourth row, second column) originates from the ability of these models to clear the central gas. While this is done by kicks in random direction in \simname{tng}, shocking the ICM and leaving behind a hot, long cooling time post-shock gas, the key in the jet model is the cocoon of jetted material present in the center, expanding outward perpendicular to the jet direction to adjust to the jet-induced over-pressure. This increases the entropy of the low-cooling time gas and moves it to larger radii and lower pressure environments, thereby increasing its cooling time. Importantly, this material is not present if the jet is heavier (i.e. having a higher momentum flux at fixed energy) because in this case, the jet is able to drill through the ICM and consequently very little jetted material remains in the centre, leading to a less efficient delay of the cooling flow \citep[qualitatively consistent with the scaling of the cocoon width described in][]{su21}. A wider opening angle (\simname{rw wind d 2}) leads to an earlier stopping, but does not fully overcome the effect. It should be noted, however, that the jet density is not the only factor of importance. In particular, simple analytic considerations of momentum density balance in the jet head rest-frame \citep[see e.g.][]{begelman84} yield a speed at which the jet head advances ($v_\rmn{a}$) of
\begin{align}
    v_\rmn{a} &= (1-(\rho_\rmn{e}/\rho_\mathrm{jet})^{1/2})^{-1} v_\mathrm{jet} \approx \left(\frac{\rho_\mathrm{jet}}{\rho_\rmn{e}}\right)^{1/2} v_\mathrm{jet}\nonumber\\ &\approx \left(\frac{ 2 \dot{E}}{A \rho_\rmn{e}}\right)^{1/3} \left(\frac{\rho_\mathrm{jet}}{\rho_\rmn{e}}\right)^{1/6}, \label{eq:advance}
\end{align}
where we assume that the external density, $\rho_\rmn{e} \gg \rho_\mathrm{jet}$ and use equation~(\ref{eq:jet_edot}) to replace $v_\mathrm{jet}$ with the jet kinetic luminosity $\dot{E}$. Equation~(\ref{eq:advance}) highlights the importance not just of the jet density, but also the external density $\rho_\rmn{e}$, jet kinematic luminosity $\dot{E}$ and cross-section $A$ to influence the behavior yet they are left constant in this study. A multi-phase external medium and entrainment further complicate the picture \citep{mukherjee16}.
The weak dependence on the jet head advance on jet density in equation~(\ref{eq:advance}), however, does not imply a weak dependence on heating of the surrounding ICM. The lateral extent of the cavity does depend more strongly on jet density \citep[see e.g.][their equation (10) for a simple model]{su21}. Related to this effect, we find the bow-shock Mach number in directions perpendicular to the jet to be increased for lighter jets, leading to increased shock-heating of the ICM in this direction. For a more detailed discussion on how jet parameters impact cocoon formation, see \citet{Su23}.
Having a look at the longer term effect of the cooling over free-fall time distribution (Fig.~\ref{fig:tcool_tff_profile_model}) there is again a strikingly universal increase in cold gas at larger radii for the models that are able to reduce the cooling times most efficiently.

We conclude from this part that the choice of parameters of the feedback model has a larger impact than resolution effects for this setup. Interestingly, very light jets have a similarly strong effect on cooling times after $50$~Myr as the kinetic feedback mode in the IllustrisTNG simulation, while models with heavier jets or directional winds are less efficient due to their reduced ability to affect the innermost ICM gas.

\subsection{Differences due to hydrodynamics code and implementation}
\label{sec:code}

The large differences for runs with different models and parameters we studied in the last subsection prompts the question of how comparable runs with independent codes are, as they employ different hydrodynamics solvers, different cooling functions, as well as a different parametrization and implementation of the feedback source terms.
To study this, we compare the simulations similar to the ones presented so far to equivalent (see Section~\ref{subsec:kys_jet_model}) ones employing the \citet{su21} jet model (\simname{kys jet b} and \simname{kys jet c}) which uses the Gizmo code, a meshless finite mass hydrodynamics solver and the FIRE-2  model for cooling. Note that the model parameter variations in this subsection are less extreme than in the previous one to allow a comparison across implementations.

Figure~\ref{fig:tcool_code} shows the cooling time distributions for different codes, each one with two parameters controlling for different momentum fluxes (model \simname{b} with lower jet momentum flux, i.e. lower jet density and higher velocity, model \simname{c} with higher momentum flux). The reference run without jet (dashed black line) differs at the low cooling time end. This can be explained by the fact that the cooling model in the Arepo simulation cuts off at $10^4$~K, thereby de facto imposing a lower limit to the cooling time (at this point, the star formation model takes over). The FIRE-2  model follows the thermal instability explicitly to lower temperatures, thus populating the low cooling-time distribution function. We speculate that the noise in the distribution function is due to details of the numerical implementation of the cooling and will not be considered further in this work. For the simulations including jets, two features are noteworthy: first, the two codes produce different cooling time distributions, the \simname{kys} implementation being less able to delay the cooling flow in this setup. Second, parameters \simname{b} creates a larger delay in cooling time compared to parameters \simname{c} for both models.

In order to explore the origin of the differences between the different codes, it is again useful to examine the phase diagram of the cooling gas. Fig.~\ref{fig:phase_diagram_code} shows this phase diagram, this time not only after $50$~Myr as in previous plots (bottom), but also while the jet is active (top after $12.5$~Myr, center after $25$~Myr). The overall response of the gas to an active jet after $12.5$~Myr is an increase in pressure (shift to the top right), with an accompanying increase in entropy (shift to the top left). The former seems to be universal across codes and parameters while the latter is more model and parameter dependent. Once the halo gas has time to hydrodynamically respond to the changed conditions, the gas expands adiabatically, decreasing its pressure while changes in entropy are more and more driven by cooling.

The large differences in the cooling time distribution function after $50$~Myr seems to be gradually developing, with differences at earlier times being much smaller. This indicates that the substantial differences in outcome between codes originates from diverging evolution of the cooling flow given small differences induced by the different jet models and by differences in the cooling functions. The response of the gas to a jet with fixed energy is qualitatively similar, though a lower efficiency of the \simname{kys jet} implementation (and of the lower density jet model \simname{c} relative to model \simname{b} at fixed implementation) is already visible after $12.5$~Myr. This implies that both, model parameter choices as well as the specific implementation are sources of uncertainty which can, due to the non-linearity of the developing cooling flow, lead to divergent results at fixed energy injection.

The cooling time maps of the simulations with different codes are shown in Fig.~\ref{fig:tcool_slices_code}. In terms of jet propagation, the two models look remarkably similar. The \simname{rw jet} models have more pronounced bow-shocks (visible in the cooling time map), indicating that the difference in entropy change could be due to differences in shock heating. This could be caused by details of the jet injection algorithm or due to the different hydrodynamics solver. Toward later times, after $250$~Myr, the \simname{rw jet} simulations, i.e. the ones run with Arepo, show a stronger tendency to shatter the jet-inflated cavity (an effect that would be suppressed if magnetic fields were present). Trends in cooling over free-fall time are similar to the ones discussed in section~\ref{sec:modelvar} and shown in Appendix~\ref{app:tc_tff}.

Overall, we conclude that differences in code, cooling function and model parameters far exceed resolution effects and uncertainties due to numerical (non-)convergence. This, in combination with the code independent trend of heavy/kinetic vs. light/thermal jets, implies that jet modeling needs to be improved in two regimes to increase the reliability of AGN feedback modeling in galaxy cluster simulations: first, a more detailed knowledge about the physical conditions in jets and second, constraints from smaller scale jet simulations about the effect at injection scales in order to establish a `correct' way to inject a jet in lower resolution galaxy cluster simulations. Finally, it is important to note that the presented setup of a fixed energy output, while instructive, is artificial since simulations of galaxy cluster evolution require a self-regulated accretion-feedback setup. This self-regulation will likely act as an attractor towards a steady-state solution with global heating-cooling balance, thereby reducing the differences discussed in this section.

\section{Self-regulated runs}
\label{sec:self_regulated}

\begin{figure}
    \centering
    \includegraphics{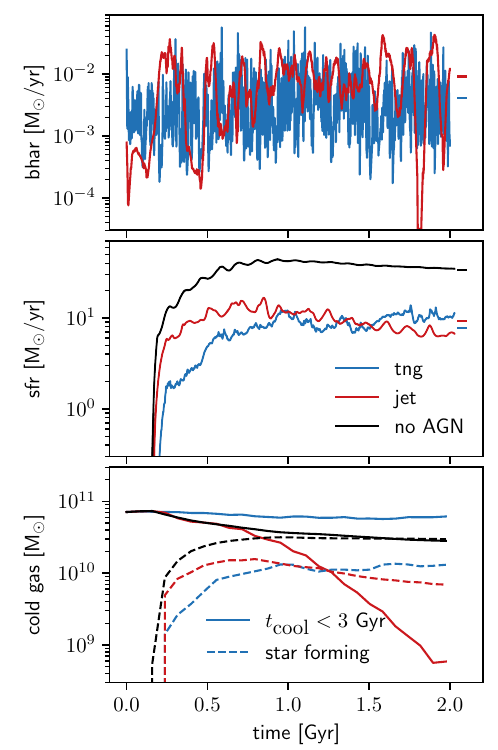}
    \caption{Black hole accretion rate, star formation rate and cooling (solid) and star forming (dashed) gas mass as a function of time for different AGN feedback models run with the Arepo code. While the \simname{tng} model acts on the star forming gas and the immediate surroundings, the \simname{jet} model acts on the cooling, not yet star forming gas, leading to a different time variability.}
    \label{fig:sfr_bhar_time}
\end{figure}

\begin{figure}
    \centering
    \includegraphics{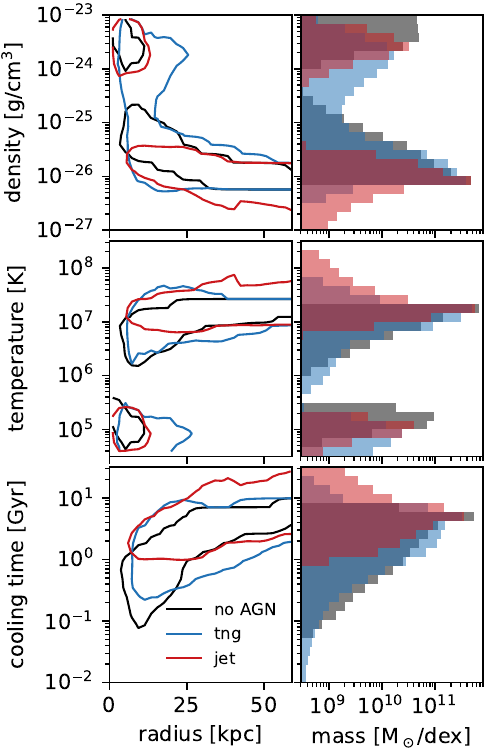}
    \caption{Gas density, temperature and cooling time as a function of radius after about 1.18 Gyr. The \simname{jet} model clearly reduces the density of the hot, dilute component more than the \simname{tng} model, leading to longer cooling times, while leaving a dense, star forming component in place. The contour is located at values of $10^8$~M$_\odot$~kpc$^{-1}$~dex$^{-1}$.}
    \label{fig:profiles}
\end{figure}

\begin{figure*}
    \centering
    \includegraphics{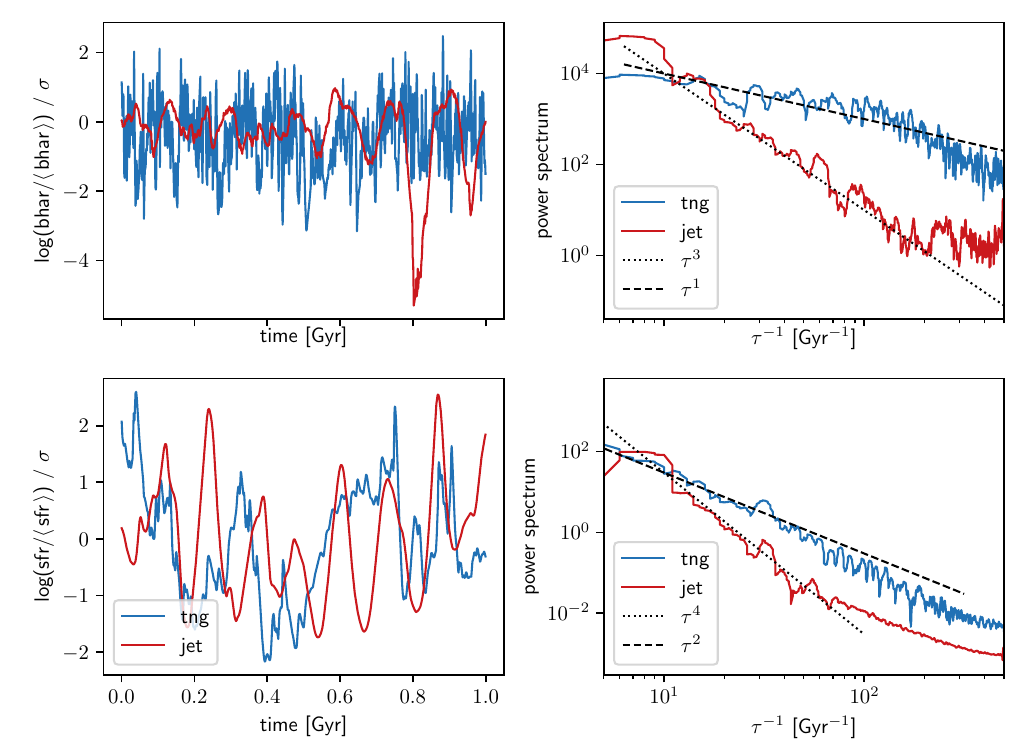}
    \caption{Left: Normalized black hole accretion rate (top) and star formation rate (bottom) of self-regulated runs using \simname{tng} and \simname{jet} feedback. The rates are extracted from 1 to 2 Gyr in the simulations, a linear fit to the respective quantity (log(sfr), log(bhar)) is subtracted, and the resulting rate is normalized by its standard deviation. Right: Power spectra of the respective rates. Both, the accretion rate and the star formation rate have larger high-frequency power for the \simname{tng} run, while the spectra decline more steeply for the \simname{jet} run.}
    \label{fig:power_spectrum}
\end{figure*}

Having established the ability of jets to delay developing cooling flows, and their resolution convergence requirements, we now focus on the more complex case of a self-regulated setup, where the jet kinetic luminosity $\dot{E}$ is coupled to the accretion rate via a fixed efficiency:
\begin{align}
    \dot{E} &= \epsilon_\text{f,kin} \, \dot{m}_\text{acc} \, c^2, \qquad
    \epsilon_\text{f,kin} = 0.1.
\end{align}
The accretion rate $\dot{m}_\text{acc}$ is estimated using the Bondi-formula and $c$ denotes the speed of light. For a self-regulated cooling-heating equilibrium state to be reached, the simulated time needs to be extended to several times the central cooling time. We perform the simulations for $2$~Gyr as a compromise, allowing an equilibrium situation to be established, yet the assumption of an isolated halo, i.e. the omission of cosmological accretion, is still a reasonable approximation of the ICM in a galaxy cluster. We compare 3 simulations in this section: a run without AGN feedback as a reference for a cooling-flow, a run with the \simname{rw jet} model and a run with the \simname{tng} kinetic feedback model. Note that we restrict the comparison in this section to simulations run with Arepo, and the jet parameters to model \simname{a}. A comparison to denser jets in a self-regulated setup is presented in \citet{ehlert22}. The black hole masses in the latter two simulations are chosen such that the resulting equilibrium star formation rates are similar, and can be considered free parameters in this setup. In particular, we note that the black hole mass in the \simname{tng} model is $3\times 10^9$~M$_\odot$, while it is only $3 \times 10^8$~M$_\odot$ for the \simname{jet} model. This has the practical consequence that the accretion rate of the \simname{tng} simulation is boosted by a factor of $100$ compared to the \simname{jet} simulation, assuming the same surrounding gas density and sound speed. If we were to run the \simname{tng} simulation with $3 \times 10^8$~M$_\odot$, the feedback would keep the black hole accretion rate at a relatively moderate rate, however, it would not be able to suppress the overall star formation rate over the entire $2$~Gyr run time. With the different masses in place, we are able to obtain simulations with comparable levels of star formation and black hole accretion, but necessarily with very different central gas densities and sound speeds. The origin of this behaviour will become clearer once we discuss the runs in more detail.

We start by showing the time evolution of the black hole accretion rate, the star formation rate and the mass in star-forming and rapidly cooling (cooling time $t_\text{c} < 3$~Gyr, but not yet star-forming) gas in Fig.~\ref{fig:sfr_bhar_time}. First, as seen in the middle panel, the \simname{tng} run shows almost an order of magnitude reduced star formation relative to the run without AGN feedback. Interestingly, the rate of star formation is reduced by a similar degree in the \simname{jet} run, yet with somewhat different time evolution. We note that the level of agreement (time averages as horizontal lines at 2~Gyr) is secondary since we have chosen different black hole masses in the two simulations. It is remarkable that light jets can mediate the cooling flow to similar levels of star formations as \simname{tng}. The black hole accretion rate shown in the top panel is comparable between the two models, with only a slightly higher overall accretion rate for the jet model. This implies that the total injected energy via AGN feedback is comparable between the two runs as well.

The bottom panel of Fig.~\ref{fig:sfr_bhar_time} shows the rapidly cooling (solid) and cold, star forming (dashed) gas components in the different simulations. While the star forming gas closely follows the star formation rate, which is a direct consequence of the employed \citet{springel03} multi-phase ISM sub-grid model, the cooling, but not yet star forming gas masses differ drastically: the gas mass in rapidly cooling gas in the \simname{no AGN} run decreases likely due to transitioning to the star-forming gas phase and ultimately forming stars. Surprisingly, the \simname{tng} run is able to retain more mass in rapidly cooling gas. The \simname{jet} simulation, on the other hand, initially follows the \simname{no AGN} run, builds up a star forming gas reservoir that is a factor of a few larger than in \simname{tng} (but still lower than the \simname{no AGN} case) and from $0.75$~Gyr onward slowly depletes the star-forming phase. More prominently, it also drastically depletes the rapidly cooling phase, in the end by two orders of magnitude. This implies, while the run applying the \simname{tng} model appears to reach a steady-state, the run using AGN driven jets undergoes a transformation on Gyr timescales (while being self-regulated in terms of smaller scale star formation and black hole accretion rate). In summary, we have two self-regulated isolated galaxy cluster simulations employing different AGN feedback models, injecting similar amounts of energy into the system, leading to a similar overall star formation suppression, yet a completely different evolution of the rapidly cooling gas, indicating that the mechanism by which they suppress star formation is fundamentally different.

To better understand how the feedback models act differently on the gas, we show the gas density, temperature and cooling time as a function of radius in 2d histograms in Fig.~\ref{fig:profiles}, left panel, and the respective distribution functions in the right panels.
We choose to show the distributions at a time well into the simulations, after the runs start to visibly diverge.
The density distributions (top panels) clearly show a bimodality of dense, central (and by definition) star forming gas and lower density halo gas.
While the bimodality exists in both simulations, the run of the \simname{jet} model completely lacks gas in between the two peaks.
By contrast, there is gas in between the two density peaks for the \simname{tng} run.
A similar trend is seen in temperature (central panel), where the hot gas corresponds to the halo, and a tail towards lower temperatures is only present in the \simname{tng} run.
The combination of the higher density and lower temperature of this `transition' phase leads to gas with substantially reduced cooling times (bottom panel) in the \simname{tng} simulations.
Note that in this panel, the star forming gas is not present since this quantity loses its meaning once the gas is in the star forming phase; i.e.\ on an effective equation of state.
We showed that the jet model is able to deplete the `transition gas', which is able to rapidly cool and join the star forming phase.
While this undoubtedly leads to a reduction of star formation, we cannot exclude a direct impact of the jet on the star forming phase, however to a lesser degree than in the IllustrisTNG model.

Finally, we focus on one of the consequences of the different ways the feedback models act on the halo gas. In particular, we consider the time variability of the black hole accretion rate and the star formation rate as shown in Fig.~\ref{fig:sfr_bhar_time}. By eye, it is already clear that the \simname{tng} run shows a significantly higher degree of high-frequency time variability than the \simname{jet} run. This is most clearly visible in the black hole accretion rate. To quantify the variability timescales, we use the logarithmic black hole accretion and star formation rates in the second half of the simulation time ($1$~Gyr to $2$~Gyr), subtract a linear fit\footnote{This is done to remove potential effects originating from star-forming gas depletion over Gyr timescales which are unrealistic due to our isolated setup.} and normalize their amplitude to further analyze the signal (Fig.~\ref{fig:power_spectrum}, left). We then calculate its power spectrum and overplot roughly matching power-law scalings in Fig.~\ref{fig:power_spectrum}, right. While the power of the black hole accretion rate scales with the inverse frequency in the \simname{tng} run, the jet run shows a $\tau^{3}$ scaling, i.e.\ the accretion rate is by far dominated by a larger variability timescale up to a timescale of $\sim100$~Myr. It is reasonable to assume that the variability on the larger timescales is more connected to hot halo properties, while more power on smaller timescales is connected to the ISM, making this result qualitatively consistent with the notion that the \simname{tng} model acts more on the ISM, while the \simname{jet} acts predominantly on the hot halo.
Interestingly, this trend also carries over to the time-variability signal in the star formation rate in these halos, presenting a potential avenue to distinguish the two models \citep[see e.g.][]{tacchella20,iyer20}. The power spectrum of the star formation rate falls off more steeply than that of the accretion rate, indicating a lower degree of variability on small time scales, likely due to star formation happening on larger spatial scales and being only indirectly (via the injected feedback energy) coupled to the black hole accretion rate.

\section{Discussion and Conclusion}
\label{sec:conclusion}

In this paper, we examine simulations of hydrodynamical jets propagating into an isolated, radiatively cooling halo. For simulations that couple jet power to the accretion rate we find:
\begin{itemize}
    \item Light hydrodynamic jets can suppress cooling flows. This differs from some early numerical studies \citep{vernaleo06, cattaneo07}, but more in line with more recent, higher resolution studies \citep[e.g.][]{li17, beckmann19}. Unlike these studies, in our simulations neither jet precession or an opening-angle are required, likely due to higher jet velocities/lower momentum fluxes/lower mass loading at fixed jet kinetic luminosity \citep{omma04,dubois10,li14, ehlert22}.
    \item Unlike the kinetic wind feedback in IllustrisTNG, collimated low-density jets predominantly act on the hot phase, reducing the amount of gas that is about to cool, while leaving the cooler, star forming gas phase intact. This difference in feedback explains the need for different `normalisations' in the Bondi-formula (i.e. different black hole masses or efficiency factors) for the two models to maintain a self-regulated equilibrium state.
    \item The different ways feedback acts leads to clear differences in the time variability in both black hole accretion rate and star formation rate. Jet feedback tends to lead to a time variability scale of $100$~Myr while wind feedback has a substantial time variability on even shorter timescales. Note that the variability in the (light) jet case is increased compared to self-regulated simulations with heavier jets \citep[see][]{gaspari11, ehlert22}.
\end{itemize}

Using simulations that cover only the onset of the cooling-flow and fixed jet kinetic luminosity, we show:
\begin{itemize}
    \item The delay of the cooling flow on $50$~Myr timescales is mostly achieved by heating and increasing the entropy of the densest gas of the hot atmosphere located in the very center, thereby reducing the cooling luminosity.
    \item Light jets achieve this via their expanding cocoon perpendicular to the jet propagation direction, pushing the ICM gas to larger radii and lower pressure environments. This is consistent with \citet{cielo14, cielo17}, finding that backflows of jetted material have a substantial impact on the gas surrounding the black hole already after several Myr. While the entropy of the ICM gas increases, the system reacts to the jet-induced overpressure quickly by expanding and the temperature returns the same levels it started out within a fraction of the central cooling time.
    \item The response of the halo to a fixed-luminosity jet is similar for different numerical methods (Arepo vs. Gizmo) and jet injection implementations (\citealt{weinberger17b} vs. \citealt{su21}). Yet, small differences in jet and cooling function can be amplified and lead to diverging results over time.
    \item Heavier jets tend to propagate outwards faster, diminishing their effect on gas in the very center. Hence, they have a smaller impact on the cooling time distribution of the most rapidly cooling gas. We find this trend independent of code and implementation details.
    \item The more efficient feedback models tend to produce more thermally unstable gas ($t_\text{c}/t_\text{ff} < 10$) at large radii ($>30$~kpc). The reason for this might be twofold: first due to outflow of low-entropy gas \citep{chen19}. Second, due to a prevention of radial inflow of cooling gas: in the pure cooling flow case, the pressure support disappears first in the center, allowing a radial inflow on timescales smaller than the cooling time. This leads to gas being transported to the center before the thermal instability fully develops. If the central pressure support is maintained due to AGN feedback, the thermal instability might develop at larger radii, leaving more gas with low $t_\text{c}/t_\text{ff}$ ratios at larger radii.
    \item The ability of jets to delay cooling flows is converged at moderate jet resolutions. In particular, convergence of feedback effects does not require convergence in jet propagation.
    \item Jet propagation ultimately converges provided the diameter of the jet is sufficiently resolved.
    \item Different parameters and models for the jets have varying ability in delaying cooling flows. This implies that the largest uncertainty remaining is the choice of the model and its parameters, and not numerical (i.e. resolution) limitations.
\end{itemize}

These results highlight the non-uniqueness of AGN feedback implementations for `solving' the cooling flow problem. We suggest using signatures directly related to jets and AGN feedback such as properties of X-ray cavities \citep{birzan20} to distinguish between mechanisms in future work.

One notable shortcoming of this set of simulations is the absence or overly simplified treatment of the multi-phase ISM surrounding the jet \citep{cielo18, mukherjee18, tanner22}. The presence of cold gas clumps could potentially alter the short-term variability of star formation in the jet run \citep{mandal21}. For current models it is challenging to model the ISM structure accurately yet cover a simulation time of several Gyr, but new numerical approaches \citep{weinberger22} might be able to overcome this problem in the future.

Using the results presented here, it is possible to use this predictive model of AGN jet feedback at the required resolution for numerical convergence in more realistic setups. This includes more massive analogues of local galaxy clusters in isolation \citep{ehlert22} and cosmological zoom simulations.

\section*{Acknowledgements}
The authors would like to thank the referee for the constructive comments that helped to improve this manuscript.
This is a paper from the Simulating Multiscale Astrophysics to Understand Galaxies (SMAUG) collaboration, a project intended to improve models of galaxy formation and large-scale structure by working to understand the small-scale physical processes that cannot yet be directly modeled in cosmological simulations. This work was supported by the Natural Sciences and Engineering Research Council of Canada (NSERC), funding reference \#CITA 490888-16. RW acknowledges support from the NSF via XSEDE allocation PHY210011. KE and CP acknowledge support by the European Research Council under ERC-CoG grant CRAGSMAN-646955, ERC-AdG grant PICOGAL-101019746, and DFG Research Unit FOR-5195. GLB acknowledges support from the NSF (AST-2108470, XSEDE grant MCA06N030), NASA TCAN award 80NSSC21K1053, and the Simons Foundation (grant 822237, “Learning the Universe”). CAFG was supported by NSF through grants AST-1715216, AST-2108230, and CAREER award AST-1652522; by NASA through grants 17-ATP17-0067 and 21-ATP21-0036; by STScI through grants HST-AR-16124.001-A and HST-GO-16730.016-A; by CXO through grant TM2-23005X; and by the Research Corporation for Science Advancement through a Cottrell Scholar Award. The authors gratefully acknowledge the Gauss Centre for Supercomputing e.V. (www.gauss-centre.eu) for funding this project by providing computing time on the GCS Supercomputer SuperMUC-NG at Leibniz Supercomputing Centre (www.lrz.de).

\section*{Data Availability}

The data underlying this article will be shared on reasonable request to the corresponding author.



\bibliographystyle{mnras}




\appendix

\section{Jet propagation at lower resolution}

\begin{figure}
    \centering
    \includegraphics{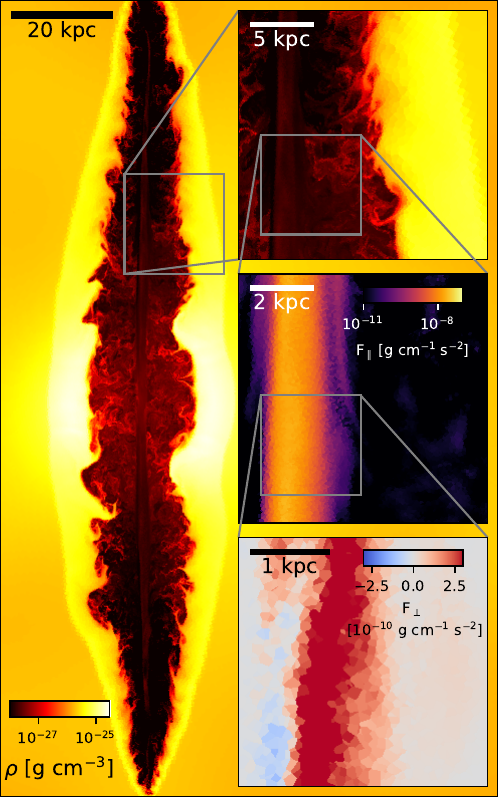}
    \caption{Same as Fig.~\ref{fig:jet_slice}, only with the slightly lower resolution simulation \textit{rw jet a 4}.}
    \label{fig:jet_slice_jet4}
\end{figure}

Fig.~\ref{fig:jet_slice_jet4} shows the \simname{rw jet a 4} simulation, which can be directly compared to Fig.~\ref{fig:jet_slice}. The most notable difference is the absence of KHI body modes in the momentum flux perpendicular to the jet propagation (bottom inlay). We speculate that the emergence of these internal effects causes the convergence in the jet position-time diagram (Fig.~\ref{fig:pos_time}).

\section{Cooling over freefall time vs radius for different codes}
\label{app:tc_tff}

\begin{figure*}
    \centering
    \includegraphics{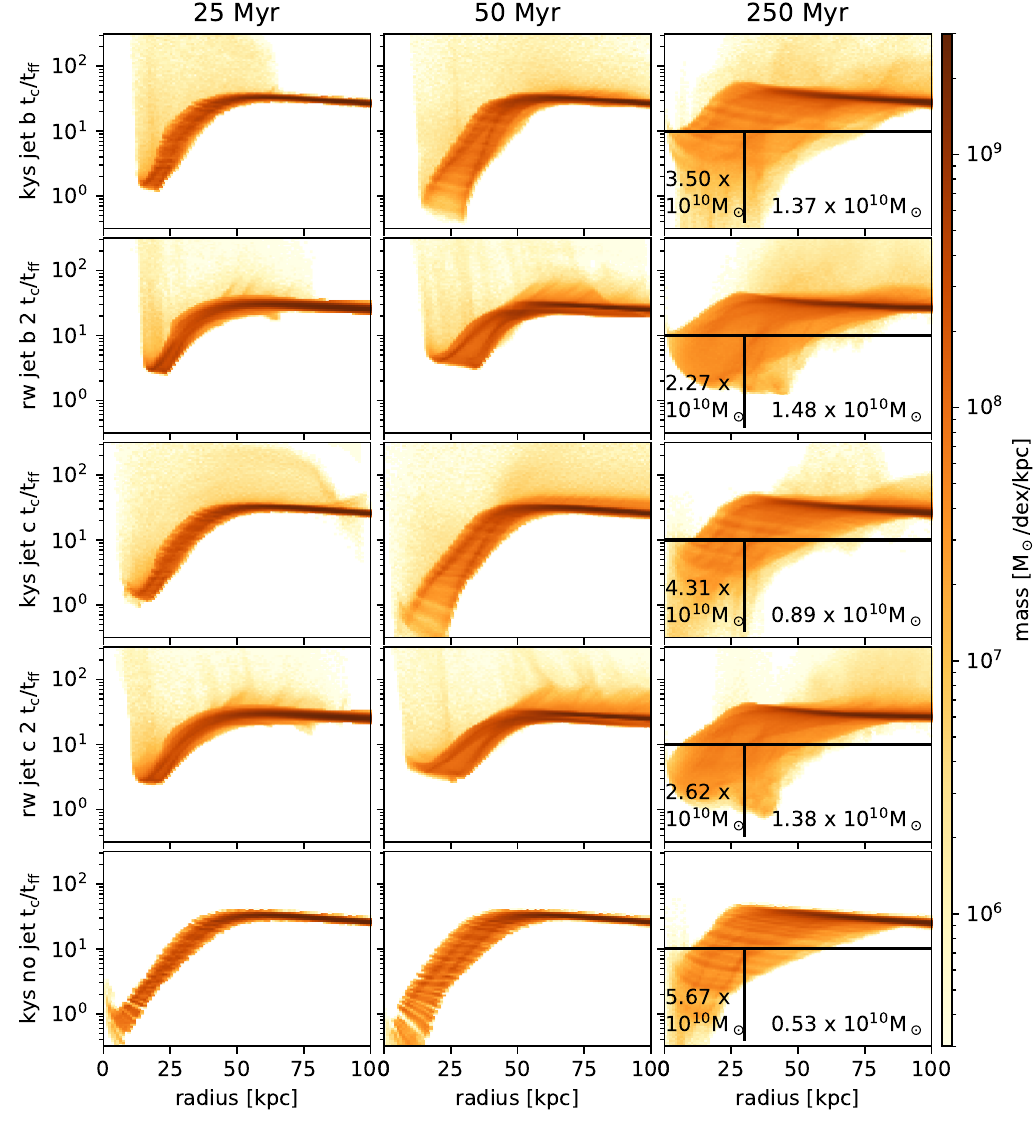}
    \caption{2d histogram of cooling over free-fall time as a function of radius.}
    \label{fig:tcool_tff_profile_code}
\end{figure*}

Figure~\ref{fig:tcool_tff_profile_code} shows the cooling over free-fall time ratio for the respective simulations. Independent of which feedback model is employed, the outburst always causes significant amounts of gas with $t_\text{c}/t_\text{ff} < 10$ at radii larger than $30$~kpc, with more efficient feedback leading to larger amounts of cooling gas at large radii.

\bsp	
\label{lastpage}
\end{document}